\documentclass[11pt]{article} 
\usepackage{graphicx}
\usepackage{bm}
\usepackage{amsmath}
\usepackage{natbib}
\usepackage{subfig}
\usepackage{amssymb}
\usepackage[normalem]{ulem} 
\setcitestyle{authoryear,citesep={; },notesep={; },round,aysep={,},yysep={;}}

\newcommand{\D}{\textrm{d}}

\begin{document}

\title{Approximating Orbits in a Rotating Gravity Field with Oblateness and Ellipticity Perturbations
}

\author{Ethan Burnett\thanks{Aerospace Engineering Sciences, University of Colorado Boulder, \texttt{ethan.burnett@colorado.edu}} \ and Hanspeter Schaub\thanks{Glenn L. Murphy Chair of Engineering, Smead Department of Aerospace Engineering Sciences, University of Colorado Boulder}%
}

\date{}

\maketitle

\begin{abstract}
This paper explores the problem of analytically approximating the orbital state for a subset of orbits in a rotating potential with oblateness and ellipticity perturbations. This is done by isolating approximate differential equations for the orbit radius and other elements. The conservation of the Jacobi integral is used to make the problem solvable to first-order in the perturbations. The solutions are characterized as small deviations from an unperturbed circular orbit.
The approximations are developed for near-circular orbits with initial mean motion $n_{0}$ around a body with rotation rate $c$. The approximations are shown to be valid for values of angular rate ratio $\Gamma = c/n_{0} > 1$, with accuracy decreasing as $\Gamma \rightarrow 1$, and singularities at and near $\Gamma = 1$.  Extensions of the methodology to eccentric orbits are discussed, with commentary on the challenges of obtaining generally valid solutions for both near-circular and eccentric orbits.
\end{abstract}

\section{Introduction}
\label{intro}
Orbital motion in uniformly rotating irregular gravity fields is generally non-integrable, greatly complicating the task of characterizing and studying the system behavior without relying on numerical simulation. Because orbits about large asteroids are often approximated (or at least qualitatively studied) by considering the dominant effects of the $C_{20} = -J_{2}$ and $C_{22}$ gravitational perturbations \citep{Scheeres}, studying motion in this particular truncated gravitational potential is a topic of significant interest for astrodynamicists and other scientists who research such bodies. Analytical developments in this problem lend useful insight for the study of orbital mechanics in the complex gravity fields of asteroids and other small bodies, which are growing targets for scientific exploration. 

By ignoring the effect of sectoral harmonics such as $C_{22}$ and only considering the zonal perturbations (valid for axisymmetric bodies in a stable minimum-energy spin state), the time-varying aspect of the gravity field is removed. This sub-problem is far simpler and more relevant to planetary orbiters and Earth satellites. As a result, there has been a lot of work in analyzing orbital motion in the axisymmetric potential of an oblate planet. Particularly noteworthy are the influential works of \cite{brouwer1} and \cite{Kozai-1959}, as well as \cite{Vinti1960}, who approximates the effects of the axisymmetric gravity field using his intermediary potential. Some more recent work is also relevant to this discussion, including work by \cite{MartinusiCDMA2015p1} and \cite{MartinusiCDMA2011p1} using an averaging technique and Brouwer-Lyddane theory to approximate the motion of low-Earth orbiting satellites, and exploiting the superintegrability of equatorial orbital dynamics under the influence of even zonal harmonics to obtain analytic expressions in terms of elliptic integrals. 

Plenty of analysis has been dedicated to the problem of motion in a second degree and order gravity field in the primary body-fixed rotating frame \citep{HuScheeresCJA}. Work has also been done to analyze this complex problem from the perspective of the perturbed osculating orbit in a non-rotating frame \citep{Scheeres, Scheeres4769Castalia1996}, and numerical studies have investigated the problem in the inertial frame \citep{HuScheeres2004}. Lastly, \cite{MahajanDelaunay} produce a Delaunay normalization of the perturbed Keplerian Hamiltonian with tesseral and sectoral harmonics. This powerful semi-analytic approach gives the variations in the equinoctial orbit elements, making use of either series expansions, iterative methods, or numerical quadrature methods to achieve the Delaunay normalization.

Explicit analytic approximations of the oblateness and ellipticity-perturbed orbit can be quite useful in some applications. For example, depending on the technique and the choice of coordinates, the process of generating time-explicit state transition matrix (STM) models for close-proximity perturbed satellite relative motion can rely heavily on finding sufficiently accurate analytic approximations of the perturbed orbit.  In \cite{Burnett2018_J2}, an analytic time-explicit STM for $J_{2}$-perturbed close-proximity satellite relative motion in near-circular orbits is obtained, leveraging an empirical approximation of the $J_{2}$-perturbed orbit radius for all inclinations. A rigorous analytic approximation is more desirable, and such a result is obtained in this paper as an exercise before the primary developments. Additionally, analytic approximations of perturbed orbit evolution can also provide some dynamical insights. 

This work is organized as follows. First, the problem of orbital motion about a rotating body with oblateness and ellipticity perturbations is reviewed. The paper considers the case of near-circular orbits about a body in a uniform minimum-energy stable spin state, and does not consider more complex cases of body spin axis precession or tumbling. The state elements considered to parameterize the orbit include polar coordinates in the osculating orbit plane. This bypasses the difficulty of using an osculating ellipse to describe a perturbed near-circular orbit \citep{Izsak1961}. Next, an analytic approximation for the orbit radius is developed using conservation of orbit energy for the $C_{20}$-only problem and conservation of the Jacobi integral for the $C_{20}$ and $C_{22}$ problem. This allows analytic solutions for the $C_{20}$ and $C_{22}$ problem to be explored to fully characterize the orbit state. The challenges of extending the presented method to orbits of arbitrary eccentricity are also considered. Applications of the model are then discussed. Finally, numerical simulations study how well the approximate solution compares to the full nonlinear trajectories.

\section{Orbits in a Rotating Gravity Field}
The dynamics problem of interest in this paper is orbital motion about a uniformly rotating body with nonzero ellipticity captured by the $C_{22}$ coefficient, and oblateness captured by the $C_{20}$ coefficient. The problem geometry is given, followed by a discussion of the analytical challenges of obtaining analytic approximations of orbital motion in this potential.
\subsection{Problem Geometry}
\begin{figure}[h!]
	\centering
	\includegraphics[width= 3.4in]{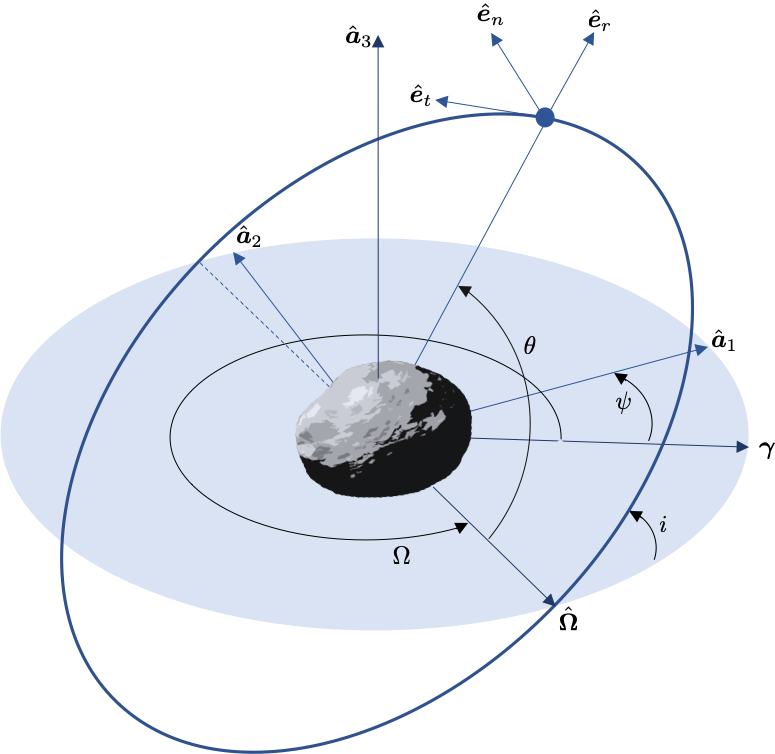}
	\caption{Problem geometry. The primary body rotates with angle $\psi$ and constant angular rate $c$, while the orbiter revolves on an osculating orbit with argument of latitude $\theta$.} 
	\label{fig:system_pg1}
\end{figure}

The relevant geometry for this dynamics problem is illustrated in Figure~\ref{fig:system_pg1}. The orbit and the rotation of the primary body-fixed frame are measured from the inertially fixed direction $\bm{\gamma}$. The primary body rotates with angle $\psi$ and constant angular rate $c$, while the orbiter revolves with argument of latitude $\theta$ and time-varying orbit-normal angular rate $\omega_{n}$, with perturbation-induced variation in $\Omega$ and $i$ as well. The main orbit equation of interest for this work is the motion of a spacecraft in orbit in a rotating primary body-fixed second degree and order gravity field \citep{Scheeres}:
 \begin{equation}
 \label{EOM_orbiter_ch6_1}
 \ddot{\bm{r}} = \bm{a}_{C_{00}} + \bm{a}_{C_{20}} + \bm{a}_{C_{22}}
 \end{equation}
\begin{equation}
\label{a_C00_ch6_1}
\bm{a}_{C_{00}} = -\mu \frac{\bm{r}}{r^{3}}
\end{equation}
\begin{equation}
\label{a_C20_ch6_1}
\bm{a}_{C_{20}} = \frac{3\mu C_{20} R^{2}}{2r^{4}}\left[\left(1-5\left(\hat{\bm{e}}_{r}\cdot \hat{\bm{a}}_{3}\right)^{2}\right)\hat{\bm{e}}_{r} + 2\left(\hat{\bm{e}}_{r}\cdot \hat{\bm{a}}_{3}\right)\hat{\bm{a}}_{3}\right]
\end{equation}
\begin{equation}
\label{a_C22_ch6_1}
\bm{a}_{C_{22}} = \frac{3\mu C_{22} R^{2}}{r^{4}}\left[-5\left(\left(\hat{\bm{e}}_{r}\cdot \hat{\bm{a}}_{1}\right)^{2}-\left(\hat{\bm{e}}_{r}\cdot \hat{\bm{a}}_{2}\right)^{2}\right)\hat{\bm{e}}_{r} + 2\left(\hat{\bm{e}}_{r}\cdot \hat{\bm{a}}_{1}\right)\hat{\bm{a}}_{1}-2\left(\hat{\bm{e}}_{r}\cdot \hat{\bm{a}}_{2}\right)\hat{\bm{a}}_{2}\right]
\end{equation}
where $\bm{r}$ is the radial vector to the orbiting satellite and $\hat{\bm{a}}_{1}$ and $\hat{\bm{a}}_{2}$ are aligned with the minimum and intermediate principal axes of inertia for a body in a stable spin state. This analysis does not consider more complicated spin states. $C_{ij}$ are coefficients from a spherical harmonics series expansion of the gravitational potential, and $C_{20} = -J_{2}$ is due to the oblateness of the body while $C_{22}$ is due to the ellipticity. These two second degree and order coefficients can be up to $\mathcal{O}(10^{-2})$ for some celestial bodies. For asteroids, they are typically the dominant secondary values in the gravitational potential after the bulk mass $C_{00} \equiv 1$ contribution. 

Instead of describing the state of the orbit using the orbit position $\bm{r}$ and velocity $\dot{\bm{r}}$ , the orbit can be described with the set of classical orbit elements, $a, e, i, \omega, \Omega, f$ (semimajor axis, eccentricity, inclination, argument of periapsis, right ascension of the ascending node, true anomaly), where $\theta = \omega + f$ is the argument of latitude. For unperturbed circular orbits, this orbit angle varies with the mean motion $n = \sqrt{\mu/a^{3}}$. Of course, under the presence of orbit perturbations, all of the orbit elements are generally time-varying. 
\subsection{Analytic Challenges}
This paper is focused on two major challenges of approximating the orbital behavior in the given situation. The first challenge is the general increase in complexity from the zonal problem introduced by the rotating gravity field. The time-varying nature of the potential results in the classical orbit energy term $E = \frac{1}{2}v^{2} - U(r)$ not being conserved. Only the more complex Jacobi integral is conserved. Furthermore, the disparate behaviors of the perturbed orbit and the uniform rotation of the body also lead to analytical difficulty. Namely, the problem dynamics are determined by the orbit plane configuration, the primary body orientation angle $\psi$ (which is linear in time in this analysis), and by an orbit latitude or anomaly which is generally quite nonlinear in time for eccentric orbits, and is itself affected by perturbations of the rotating body. To analytically describe solution behavior in terms of initial conditions and a single independent variable, the different angular behaviors must be reconciled. 

For cases with a consistently near-circular orbit, additional challenges are introduced. In such cases, the behavior of the argument of latitude is approximately linear in time, $\theta \approx \theta_{0} + nt$, but the classical orbit description in terms of an osculating ellipse becomes less useful. In particular, the classical elements of eccentricity, argument of periapsis, and true anomaly oscillate rapidly in a manner that cannot be approximated by considering small variations about a mean or initial value. The eccentricity also appears as a small divisor in the formulas for the argument of periapsis and true or mean anomalies. These difficulties can be removed by considering alternate element formulations in terms of the troublesome classical elements, such as the equinoctial elements \citep{broucke2} or Poincar\'e canonical elements. By the same manner, these alternate elements are not subject to large and rapid variations due to perturbations, and are thus potentially more suitable as a choice of coordinates in these situations than the classical elements. However, the variational equations for the equinoctial elements must typically be numerically integrated using Kepler's equation at each integration step \citep{broucke2}, an unfortunate property of the independent variable for the purposes of this work. For the perturbed near-circular orbit problem, one would expect more mathematical simplicity instead of complexity \citep{Izsak1961}. A concise analytic approximation demands a simpler choice of coordinates. The approach in this paper uses the classical elements $\Omega$ and $i$ to describe the orientation of the perturbed orbit plane, and polar coordinates ($r$, $\theta$, $\dot{r}$, $\omega_{n}$) to parameterize the remaining state elements. This treatment avoids any direct use of the eccentricity, argument of periapsis, or true anomaly, and any associated difficulties of using elements explicitly derived from these. It also remains well-defined for equatorial orbits simply by redefining $\theta$ as the rotation from $\hat{\bm{\gamma}}$ instead of the undefined ascending node. While particularly well-suited for the small eccentricity problem, this approach could also inspire extensions to solutions for more eccentric orbits.


\section{Expressions for the Perturbed Orbit Radius}
In this paper, let $\omega_{n}$ denote the angular rate around the orbit, $\omega_{n} = \bm{\omega}_{O/N}\cdot\hat{\bm{e}}_{n}$. This is related to the argument of latitude rate by $\dot{\theta} = \omega_{n} - \dot{\Omega}\cos{i}$, where the second term is due to the deviation and regression of the node from which $\theta$ is measured \citep{PrussingConway2013}. The angular rate $\omega_{n}$ is one of the six chosen state quantities, while the argument of latitude rate $\dot{\theta}$ is not. Note that the velocity of the orbiter in the orbit plane is defined as $v = \sqrt{\dot{r}^{2} + r^{2}\omega_{n}^{2}}$, because the $\hat{\bm{e}}_{r}$ component of $\bm{\omega}_{O}$ does not contribute to the orbital velocity. 
%

Using Eqs. \eqref{a_C20_ch6_1} - \eqref{a_C22_ch6_1}, the angular rates $\dot{\Omega}$ and $\dot{i}$ can be computed, then shown to be of the following scale or smaller, assuming $|C_{20}| \geq |C_{22}|$:
\begin{equation}
\label{ScaleOErate1}
\dot{\Omega}, \dot{i} = \mathcal{O}\left(C_{20}\left(\frac{R}{r}\right)^{2}\left(\frac{1}{\rho^{3}}\right)\frac{n^{2}}{\omega_{n}}\right)
\end{equation}
where $\rho = r/a$, $\omega_{n} = \mathcal{O}(n)$, and it is assumed $|C_{20}(R/r)^{2}| \ll 1$. In orbits for which this assumption holds, if the eccentricity is reasonably small, then $|\omega_{n}| \gg |\dot{\Omega}|, |\dot{i}|$ also holds.
%



The orbit radius approximation approach begins by expressing the radial part of Newton's second law in spherical coordinates:
\begin{equation}
\label{rddotvec3}
\ddot{r} - \omega_{n}^{2}r = -\frac{\mu}{r^{2}} + R_{C_{20}} + R_{C_{22}}
\end{equation}
where  $R_{C_{ij}} = \bm{a}_{C_{ij}}\cdot\hat{\bm{e}}_{r}$.
The following time-varying differential equation is obtained for the orbit radius by substituting the radial components of the disturbance accelerations:
\begin{equation}
\label{rddot1}
\begin{split}
\ddot{r} - \omega_{n}^{2}r =  -\frac{\mu}{r^{2}} + \frac{\mu}{r^{4}}\bigg(\frac{3}{2}C_{20}&R^{2}\left(1 - 3\sin^{2}{i}\sin^{2}{\theta}\right) \\ + & \ 3C_{22}R^{2}\bigg(3\sin{\left(2\left(\Omega - \psi\right)\right)}\cos{i}\sin{2\theta}  \\ - & \ \frac{3}{4}\cos{\left(2\left(\Omega-\psi\right)\right)}\left(1 + 3\cos{2\theta} - 2\cos{2i}\sin^{2}{\theta}\right)\bigg)\bigg)
\end{split}
\end{equation} 

To isolate the dynamics of the orbit radius, the currently unknown $\omega_{n}$ term in Eq.~\eqref{rddot1} must be rewritten. This is done by isolating the $\omega_{n}$ terms in an integral of motion and re-arranging to obtain an expression for $\omega_{n}$ that is a function of only $r$, $\dot{r}$, $\theta$, $\Omega$, $i$, $\psi$, and the conserved value of that integral of motion. In the expression for $\omega_{n}$, functions of quantities $\Omega$ and $i$ primarily appear pre-multiplied by small parameters that are functions of $C_{20}$ and $C_{22}$. Then, the assumption that $\dot{\Omega}$ and $\dot{i}$ are ``small" equivalently results in using the initial values $\Omega_{0}$ and $i_{0}$ in these terms. In the following sections, this equation will be solved using conservation of energy for the case $C_{22} = 0$ for all inclinations, and it will be solved for an expression accurate for inclinations below a critical value for $C_{22} \neq 0$ using conservation of the Jacobi integral.

\subsection{Approximate Solution Using Conservation of Energy for Near-Circular Orbits}
To introduce the perturbative procedure for approximating the orbit radius, the simpler $C_{20}$-only zonal problem is first solved. For the case of negligible influence by $C_{22}$, using the orbit energy and the substitutions $r(t) \approx r_{0}(1 + \xi(t))$ where $\xi \sim \mathcal{O}(1/r_{0})$ and $\dot{r} \approx r_{0}\dot{\xi}$, it is possible to approximate the behavior of Eq.~\eqref{rddot1} with a simpler differential equation that is an explicit function of time. 
Below, the total orbit energy (only conserved when either $C_{22}$ or $c$ are zero) is given, where $U(r)$ is the gravitational potential:
\begin{equation}
\label{OrbitEnergy1}
E = \frac{1}{2}v^{2} - U(r) 
\end{equation}
\begin{equation}
\label{OrbitPotential}
\begin{split}
U(\bm{r},t) = & \ \frac{\mu}{r} + \frac{\mu}{r^{3}}\Bigg[C_{20}R^{2}\left(\frac{3}{4}\sin^{2}{i}\left(1 - \cos{2\theta}\right) - \frac{1}{2}\right) \\ & + 3C_{22}R^{2}\bigg(\frac{1}{2}\sin^{2}{i}\cos{\left(2(\Omega - \psi)\right)} + \cos^{4}{\left(\frac{i}{2}\right)}\cos{\left(2(\Omega + \theta - \psi)\right)} \\ & + \sin^{4}{\left(\frac{i}{2}\right)}\cos{\left(2(\Omega - \theta - \psi)\right)}\bigg)\Bigg]
\end{split}
\end{equation} 
The orbit energy is written in terms of $\omega_{n}$:
\begin{equation}
\label{OrbitEnergy1}
E = \frac{1}{2}\left(r^{2}\omega_{n}^{2} + \dot{r}^{2}\right) - U(r) 
\end{equation}
When $C_{22} = 0$, $E = E_{0} \ \forall t$ and the following may be written:
\begin{equation}
\label{thetadotsqr1}
\omega_{n}^{2} = \frac{2\left(E_{0} + \tilde{U}(r)\right) - \dot{r}^{2}}{r^{2}}
\end{equation}
where $\tilde{U}(r)$ contains only the $C_{00}$ and $C_{20}$ components of the gravitational potential.

Using the substitution $r(t) = r_{0}(1 + \xi(t))$, and substituting $\omega_{n}^{2}$ using Eqs.~\eqref{OrbitPotential} -~\eqref{thetadotsqr1}, Eq.~\eqref{rddot1} is expanded about $\xi = 0$, retaining terms linear in $\xi$:
\begin{equation}
\label{rddot1b}
\ddot{\xi} + \left(2\left(\frac{\mu}{r_{0}^{3}} + \frac{E_{0}}{r_{0}^{2}}\right) - 4\frac{\mu}{r_{0}^{3}}\overline{f}\right)\xi - \left(\frac{\mu}{r_{0}^{3}} + 2\frac{E_{0}}{r_{0}^{2}}-\frac{\mu}{r_{0}^{3}}\overline{f}\right) = 0
\end{equation}
where $\overline{f}$ is a function associated with the $C_{20}$ component of the gravitational potential:
\begin{equation}
\label{fbar1}
\begin{split}
\overline{f}(t) = & \ C_{20}\frac{R^{2}}{r_{0}^{2}}\left(\frac{3}{4}\sin^{2}{i_{0}}\left(1 - \cos{2\theta}\right) - \frac{1}{2}\right) 
\end{split}
\end{equation}
Thus, the problem has been transformed to a study of (assumed small) variations about the initial value $r(0) = r_{0}$, an approximation which significantly simplifies the problem. The following change of time variables enables the subsequent non-dimensionalization of Eq.~\eqref{rddot1b}:
\begin{equation}
\label{Ttransform1}
\tau = \left(\frac{\mu}{r_{0}^{3}}\right)^{1/2}t, \ \frac{{\text{d}}}{{\text{d}}t} =  \left(\frac{\mu}{r_{0}^{3}}\right)^{1/2}\frac{{\text{d}}}{{\text{d}}\tau}
\end{equation}
\begin{equation}
\label{rddot1c}
\begin{split}
\xi'' + & \left(2\left(1 + \frac{E_{0}r_{0}}{\mu}\right)  - 4\overline{f}\right)\xi - \left(\left(1+ 2\frac{E_{0}r_{0}}{\mu}\right) - \overline{f}\right) = 0
\end{split}
\end{equation}
Note that here we redefine $\left( \ \right)' = {\text{d}}/{\text{d}}\tau\left( \ \right)$. The change of time variables renders $\xi(\tau)$, $\xi'(\tau)$, and $\xi''(\tau)$ to all be of the same order. In this derivation, it is assumed that $\xi$ and $\overline{f}$ are both similarly small (e.g. $10^{-2}$), denoted $\mathcal{O}(\epsilon)$. This derivation could be modified to accommodate different relative scales. The smallness of $\xi$ depends on the orbit not deviating drastically from the unperturbed geometry, and the scale of $\overline{f}$ depends on the altitude and the size of $C_{20}$. 

This system can be initiated (without loss of generality) with $\theta_{0} = 0$, then substitution of $\theta \approx n_{0}t$, $n_{0} = \sqrt{\mu/a_{0}^{3}}$, renders the function $\overline{f}$ as an explicit function of time $t$. Note that non-circular orbit angular frequency variations would appear pre-multiplied by other small terms (i.e. terms involving $C_{20}$), and are thus neglected. Finally, the substitution $t = \left(\mu/r_{0}^{3}\right)^{-1/2}\tau$ renders everything a function of the dimensionless time: $\theta = \left(r_{0}^{3}/a_{0}^{3}\right)^{1/2}\tau$.

Identifying the small parameters $\xi$ and $\overline{f}$ as $\mathcal{O}(\epsilon)$, the $\mathcal{O}(\epsilon)$ part of Eq.~\eqref{rddot1c} is given below:
\begin{equation}
\label{rddot1f}
\xi'' + 2\left(1 + \frac{E_{0}r_{0}}{\mu}\right)\xi = \left(1+ 2\frac{E_{0}r_{0}}{\mu}\right)-\overline{f}
\end{equation}
To first order, $\xi$ obeys simple sinusoidal dynamics with an oscillatory forcing term due to the negative of the $C_{20}$ component of the potential. 
This first-order equation can be solved using the method of undetermined coefficients, noting that the harmonic forcing term has different frequencies from the homogeneous solution. 
The solution of Eq.~\eqref{rddot1f} is the sum of the homogeneous and particular solutions given below:
\begin{equation}
\label{xiH1}
\xi_{h}(\tau) = D\cos{\left(\sqrt{2\eta_{1}}\tau\right)} + E\sin{\left(\sqrt{2\eta_{1}}\tau\right)}
\end{equation}
\begin{equation}
\label{xiP1}
\xi_{p}(\tau) = A\cos{\left(\omega_{p}\tau\right)} + B\sin{\left(\omega_{p}\tau\right)} + C
\end{equation}
where the quantities $\eta_{1}$ and $\omega_{p}$ are given:
\begin{equation}
\label{eta1}
\eta_{1} = 1 + \frac{E_{0}r_{0}}{\mu}, \  \omega_{p} = 2n_{0}\left(\frac{\mu}{r_{0}^{3}}\right)^{-1/2}
\end{equation}
Substituting the particular solution into Eq.~\eqref{rddot1f}, the following equations are obtained in terms of the undetermined coefficients $A$, $B$, and $C$:
\begin{equation}
\label{xiPcoeff1}
\begin{split}
A\left(2\eta_{1} - \omega_{p}^{2}\right) = & \ \frac{3}{4}C_{20}\frac{R^{2}}{r_{0}^{2}}\sin^{2}{i} \\
B = & \ 0 \\
2\eta_{1}C = & \ 1 + 2\frac{E_{0}r_{0}}{\mu} + \frac{1}{2}C_{20}\frac{R^{2}}{r_{0}^{2}}\left(1 - \frac{3}{2}\sin^{2}{i}\right)
\end{split}
\end{equation}
Letting $\alpha = C_{20}(R/r_{0})^{2} \sim \mathcal{O}(\epsilon)$, the following values are obtained:
\begin{equation}
\label{xiPcoeff2}
A = \frac{3}{4}\alpha\left(\frac{\sin^{2}{i}}{2\eta_{1} - \omega_{p}^{2}}\right), \ B = 0, \ C = \frac{1}{4}\alpha\left(\frac{1 - \frac{3}{2}\sin^{2}{i}}{\eta_{1}}\right) + \frac{1}{2\eta_{1}}\left(1 + 2\frac{E_{0}r_{0}}{\mu}\right)
\end{equation}
The first initial condition is $r(0) = r_{0}\left(1 + \xi(0)\right) = r_{0}$. The next initial condition on $\xi$ is given from the following expression:
\begin{equation}
\label{rprime1}
r'(0) = \left(\frac{\mu}{r_{0}^{3}}\right)^{-1/2}\dot{r}(0) =  \left(\frac{\mu}{r_{0}^{3}}\right)^{-1/2}r_{0}\dot{\xi}(0) = r_{0}\xi'(0)
\end{equation}
\begin{equation}
\label{xiprime1}
\xi'(0) = \left(\frac{\mu}{r_{0}^{3}}\right)^{-1/2}\frac{\dot{r}_{0}}{r_{0}} = E\sqrt{2\eta_{1}}
\end{equation}
Thus, $D$ and $E$ are obtained from the initial conditions:
\begin{equation}
\label{Ecoeff1}
D = -A - C, \ E = \left(\frac{\mu}{r_{0}^{3}}\right)^{-1/2}\frac{\dot{r}_{0}}{r_{0}}\left(\frac{1}{\sqrt{2\eta_{1}}}\right)
\end{equation}
The approximate solution for $\xi(\tau)$ is given by the sum of Eqs.~\eqref{xiH1} and~\eqref{xiP1} with the coefficients given in Eqs.~\eqref{xiPcoeff2} and~\eqref{Ecoeff1}, thus approximating $r(\tau) = r_{0}\left(1 + \xi(\tau)\right)$ to first order:
\begin{equation}
\label{xiF1}
\xi(\tau) = A\cos{\left(\omega_{p}\tau\right)} + C + D\cos{\left(\sqrt{2\eta_{1}}\tau\right)} + E\sin{\left(\sqrt{2\eta_{1}}\tau\right)}
\end{equation}
This simple approximation is accurate for sufficiently small initial eccentricity ($e_{0} \sim 10^{-3}$) and for all inclinations. Accuracy is less dependent on the osculating eccentricity, which can generally grow to larger values ($10^{-2}$) at some points in the orbit.
The accuracy for small eccentricity is captured in the terms $C$, $D$, and $E$. For example, setting $\omega = f_{0} = 0$, the equation for $\xi$ reduces to:
\begin{equation}
\label{xiF1SmallE}
\xi(\tau) = C\left(1 - \cos{\left(\sqrt{2\eta_{1}}\tau\right)}\right)
\end{equation}
Then, $r(\tau) = a(1-e)\left(1 + \xi(\tau)\right)$. Making the necessary substitutions, then expanding to first-order in eccentricity, the classical first-order expansion \citep{battin1} is recovered:
\begin{equation}
\label{xiF1SmallE2}
r(t) \approx a\left(1 - e\cos{\left(n_{0}t\right)}\right)
\end{equation}

When $C_{22} \neq 0$, the procedure discussed in this sub-section can still be applied for cases where the primary body is sufficiently slowly rotating ($\dot{\psi} = c \ll n$). With the slow gravity field rotation, the orbit energy will be nearly conserved on the time scale of a single orbit, and this analysis can be extended to approximate variations in $r(t)$ for several orbits. However, this scenario is somewhat rare in nature, and a less restricted solution is sought.

\subsection{Approximate Solution Using Jacobi Integral for Near-Circular Orbits}
With the introduction of the time-varying potential terms due to $C_{22}$, energy is no longer conserved in the inertial frame for this system. However, there is still a conserved quantity that can be used, existing for general uniformly rotating gravitational potentials \citep{GalacticDynamics}. Given a general smooth and continuous orbit potential $U(\bm{r})$, the Lagrangian is given below, along with the conjugate momenta:
\begin{equation}
\label{Lagrangian1}
\mathcal{L} = \frac{1}{2}\|\bm{r}' + \bm{\omega}_{B/N}\times\bm{r}\|^{2} + U(\bm{r})
\end{equation}
\begin{equation}
\label{Momentum1}
\bm{p} = \frac{\partial\mathcal{L}}{\partial\bm{r}'} = \bm{r}' + \bm{\omega}_{B/N}\times\bm{r}
\end{equation}
where $\bm{\omega}_{B/N}$ is the angular velocity of the primary body-fixed rotating frame and $\bm{r}'$ is the velocity seen in that frame. The Hamiltonian is given below:
\begin{equation}
\label{Hamiltonian1}
H_{J} = \bm{p}\cdot\bm{r}' - \mathcal{L} = \bm{p}\cdot(\bm{p} - \bm{\omega}\times\bm{r}) - \frac{1}{2}p^{2} - U
\end{equation}
This may be written in the following form in terms of the angular momentum $\bm{L}$, and it is noted that $H_{J}$ has no explicit time dependence. Thus the derivative along any orbit ${\D}H_{J}/{\D}t$ vanishes and $H_{J}$ is thus an integral, called the Jacobi integral \citep{GalacticDynamics}:
\begin{equation}
\label{HamiltonJI1}
H_{J} = \frac{1}{2}p^{2} - U - \bm{\omega}_{B/N}\cdot(\bm{r}\times\bm{p}) = H - \bm{\omega}_{B/N}\cdot\bm{L}
\end{equation}

Using the celestial mechanics convention $\bm{p} = \bm{v}$ and $\bm{L} = \bm{h}$, Eq.~\eqref{HamiltonJI1} is adapted to the notation in this paper by writing $\bm{\omega}_{B/N} = c\hat{\bm{a}}_{3}$. Then the Jacobi integral is recognized in the following form for the second degree and order gravitational potential \citep{Scheeres, Scheeres1999CMDA}:
\begin{equation}
\label{JacobiC20C22}
J = \frac{1}{2}v^{2} - \frac{\mu}{r} - hc\cos{i} - U_{2}(\bm{r},t) = J_{0}
\end{equation}
where $U_{2}(r)$ isolates the $C_{20}$ and $C_{22}$ components of the gravitational potential given in Eq.~\eqref{OrbitPotential}, $h = r^{2}\omega_{n}$ is the angular momentum, and $c$ is the primary body rotation rate. 
Because it has been shown that this integral exists for any uniformly rotating potential, the procedure used in this paper can in principal be extended to more complex gravitational fields. 

Below, the Jacobi integral is written in terms of $\omega_{n}$:
\begin{equation}
\label{JacobiC20C22v2}
J = \frac{1}{2}\left(r^{2}\omega_{n}^{2} + \dot{r}^{2}\right) - \frac{\mu}{r} - cr^{2}\omega_{n}\cos{i} - U_{2}(\bm{r},t) = J_{0}
\end{equation}
Re-arranging equation~\eqref{JacobiC20C22v2}, an equation for $\omega_{n}$ is found in terms of $r$ and $\dot{r}$:
\begin{equation}
\label{thetadotfromJ}
\omega_{n} = c\cos{i} \pm \sqrt{c^{2}\cos^{2}{i} - \left(\frac{\dot{r}^{2}}{r^{2}} - \frac{2}{r^{2}}\left(\frac{\mu}{r} + U_{2}(\bm{r},t) + J_{0}\right)\right)}
\end{equation}
where the sign is negative for prograde orbits and positive for retrograde orbits. The equations are subsequently developed for prograde orbits. Eq. \eqref{thetadotfromJ} highlights the close relationship between $\omega_{n}$ and $c$. Note that as $c\rightarrow 0$, the Jacobi integral simply becomes the orbit energy, aligning with the expectation that the conservation of energy derivation in the previous section would also be valid for very slowly-rotating bodies with $C_{22} \neq 0$. There is one additional complication for this more general case: the small variations in the inclination must be accounted for in any term that is larger than $\mathcal{O}(\epsilon)$. In this paper, that turns out to mean any term not linear in $C_{20}$ or $C_{22}$. In particular, these are the $c\cos{i}$ terms in Eq.~\eqref{thetadotfromJ}.

To obtain an expression of sufficient accuracy for the $\mathcal{O}(\epsilon)$ derivation in this paper, one may integrate the Gauss planetary equation for inclination with the following first-order approximation:
\begin{equation}
\label{iApprox_v1}
\begin{split}
i(t^{*}) \approx i_{0} + \frac{3\mu R^{2}}{h_{0}r_{0}^{3}}\int_{0}^{t_{f}}\bigg(2C_{22}&\sin{(2(\Omega_{0}-\psi))}\cos^{2}{\theta}\sin{i_{0}} \\ & + \frac{1}{4}\left(C_{20} + 2C_{22}\cos{(2(\Omega_{0}-\psi))}\right)\sin{2\theta}\sin{2i_{0}}\bigg){\D}t
\end{split}
\end{equation}
where $\psi = ct$, by construction, from the freedom of choice in defining the arbitrary reference direction $\hat{\bm{\gamma}}$. In the case of near-circular orbits, $\theta \approx \theta_{0} + n_{0}t$, and the result is given below:
\begin{equation}
\label{iApprox_v1Solve}
\begin{split}
i(t) \approx & \ i_{0} + \frac{3\mu R^{2}}{4n_{0}h_{0}r_{0}^{3}}C_{20}\left(\sin{(\theta-\theta_{0})}\sin{(\theta+\theta_{0})}\right)\sin{2i_{0}} 
\\ & + \frac{3\mu R^{2}}{h_{0}r_{0}^{3}}C_{22}\Bigg(\frac{1}{8c(c-n_{0})(c+n_{0})}\Big(-2\Big[2(c-n_{0})(c+n_{0})\cos{2\Omega_{0}} 
\\ &+ c(c-n_{0})\cos{(2(\Omega_{0}-\theta_{0}))} + (c+n_{0})\big(c\cos{(2(\Omega_{0}+\theta_{0}))} 
\\ & + 2(n_{0}-c)\cos{(2(\Omega_{0}-\psi))} - c\cos{(2(\Omega_{0}+\theta-\psi))}\big) \\ & + c(n_{0}-c)\cos{(2(\Omega_{0}-\theta-\psi))}\Big]\sin{i_{0}} 
\\ &+ c\Big[(c-n_{0})\cos{(2(\Omega_{0}-\theta_{0}))}  -(c+n_{0})\cos{(2(\Omega_{0}+\theta_{0}))} 
\\ & + (c+n_{0})\cos{(2(\Omega_{0}+\theta-\psi))} + (n_{0}-c)\cos{(2(\Omega_{0}-\theta-\psi))}\Big]\sin{2i_{0}}\Big)\Bigg)
\end{split}
\end{equation}
Writing this as $i(t) \approx i_{0} + \delta i(t)$, for which $\delta i(t)$ is the small time-varying deviation in inclination due to the gravity field, $\delta i^{2}$ is assumed negligible in this derivation, and the $\cos{i}$ term in Eq.~\eqref{thetadotfromJ} becomes $\cos{i_{0}} - \sin{i_{0}}\delta i(t)$. 

Substituting Eq.~\eqref{thetadotfromJ} and reusing the change of variables $r(t) = r_{0}(1 + \xi(t))$ and non-dimensionalization of time  $\tau = \left(\mu/r_{0}^{3}\right)^{1/2}t$, the following dimensionless equations are obtained from Eq.~\eqref{rddot1}:
\begin{equation}
\label{rddotP21}
\xi'' - \frac{r_{0}^{3}}{\mu}\left(\frac{\omega_{n}^{2}r}{r_{0}}\right) = -\frac{1}{(1 + \xi)^{2}} - \frac{3}{(1+\xi)^{4}}\overline{f}(t)
\end{equation}
\begin{equation}
\label{rddotP2s1}
\begin{split}
\frac{r_{0}^{3}}{\mu}\left(\frac{\omega_{n}^{2}r}{r_{0}}\right) = & \ 2c^{2}\left(\frac{r_{0}^{3}}{\mu}\right)\left(\cos^{2}{i_{0}} - 2\cos{i_{0}}\sin{i_{0}}\delta i\right)\left(1 + \xi\right) \\ & - c\left(\frac{r_{0}^{3}}{\mu}\right)^{1/2}\left(\cos{i_{0}} - \sin{i_{0}}\delta i\right)\bigg[4c^{2}\left(\frac{r_{0}^{3}}{\mu}\right)\left(\cos^{2}{i_{0}} - 2\cos{i_{0}}\sin{i_{0}}\delta i\right)(1 + \xi)^{2} \\ & - \ 4\left(\xi'^{2} - 2\left(\frac{1}{1+\xi} + \frac{1}{\left(1+\xi\right)^{3}}\overline{f} + \frac{J_{0}r_{0}}{\mu}\right)\right)\bigg]^{1/2} \\ & - \bigg[\frac{\xi'^{2}}{1+\xi} - \frac{2}{1+\xi}\left(\frac{1}{1+\xi} + \frac{1}{(1+\xi)^{3}}\overline{f} + \frac{J_{0}r_{0}}{\mu}\right)\bigg]
\end{split}
\end{equation}
where $\overline{f}$ is a function associated with the second degree and order components of the gravitational potential, now including $C_{22}$:
\begin{equation}
\label{fbar1}
\begin{split}
\overline{f}(t) = & \ C_{20}\frac{R^{2}}{r_{0}^{2}}\left(\frac{3}{4}\sin^{2}{i_{0}}\left(1 - \cos{2\theta}\right) - \frac{1}{2}\right) + 3C_{22}\frac{R^{2}}{r_{0}^{2}}\bigg(\frac{1}{2}\sin^{2}{i_{0}}\cos{\left(2(\Omega_{0} - \psi)\right)} \\ &+ \cos^{4}{\left(\frac{i_{0}}{2}\right)}\cos{\left(2(\Omega_{0} + \theta - \psi)\right)} + \sin^{4}{\left(\frac{i_{0}}{2}\right)}\cos{\left(2(\Omega _{0}- \theta - \psi)\right)}\bigg)
\end{split}
\end{equation}
Noting that $\xi$ and $\overline{f}$ are $\mathcal{O}(\epsilon)$, Eq.~\eqref{rddotP2s1} is reduced to an expression that is linear in $\xi$ and $\overline{f}$. This is done by factoring and binomial expanding the square root term and the $(1 + \xi)^{k}$ terms. The final result for this term is reproduced below:
\begin{equation}
\label{rddotP2s2}
\begin{split}
\frac{r_{0}^{3}}{\mu}\left(\frac{\omega_{n}^{2}r}{r_{0}}\right) \approx & \ 2\bigg(\gamma_{1} - \frac{\sqrt{\gamma_{1}}(\gamma_{1}-1)}{(\gamma_{1} + 2\gamma_{2})^{1/2}} - \frac{J_{0}r_{0}}{\mu} - 2\bigg)\xi \\ & + 2\left(1 + \overline{f} + \frac{J_{0}r_{0}}{\mu} + \left((\gamma_{1} + 2\gamma_{2})^{1/2}\gamma_{4} - 2\gamma_{3}\right)\delta i\right) \\ & + \left(2\gamma_{1} - 2\sqrt{\gamma_{1}}\left((\gamma_{1} + 2\gamma_{2})^{1/2}\left(1 + \frac{\overline{f} - \gamma_{3}\delta i}{\gamma_{1} + 2\gamma_{2}}\right)\right)\right)
\end{split}
\end{equation}
where the $\gamma_{i}$ terms are defined below:
\begin{equation}
\label{gamma1}
\gamma_{1} = c^{2}\left(\frac{r_{0}^{3}}{\mu}\right)\cos^{2}{i}, \ \gamma_{2} = 1 + \frac{J_{0}r_{0}}{\mu}, \ \gamma_{3} = c^{2}\left(\frac{r_{0}^{3}}{\mu}\right)\cos{i_{0}}\sin{i_{0}}, \ \gamma_{4} = \gamma_{3}/\sqrt{\gamma_{1}}
\end{equation} 
The remaining terms in Eq.~\eqref{rddotP21} are more easily simplified. The $\mathcal{O}(\epsilon)$ part of Eq.~\eqref{rddotP21} is linear in $\xi$, and the final expression for the linear ODE is reproduced below:
\begin{equation}
\label{xiODEp2}
\xi'' + 2\eta_{2}\xi = 2\eta_{3} - \varphi\overline{f} + \vartheta\delta i
\end{equation}
\begin{equation}
\label{eta2}
\eta_{2} = \gamma_{2} + \sqrt{\gamma_{1}}\left(\frac{\gamma_{1} - 1}{(\gamma_{1} + 2\gamma_{2})^{1/2}}\right)-\gamma_{1}
\end{equation}
\begin{equation}
\label{eta3}
\eta_{3} =\gamma_{2} - \frac{1}{2} - \ \sqrt{\gamma_{1}(\gamma_{1} + 2\gamma_{2})} + \gamma_{1}
\end{equation}
\begin{equation}
\label{varphi_sdo}
\varphi = 1 + 2\sqrt{\gamma_{1}}\left(\frac{1}{(\gamma_{1} + 2\gamma_{2})^{1/2}}\right)
\end{equation}
\begin{equation}
\label{vartheta}
\vartheta = 2\left(\left(\sqrt{\frac{\gamma_{1}}{\gamma_{1} + 2\gamma_{2}}} - 2\right)\gamma_{3} + (\gamma_{1} + 2\gamma_{2})^{1/2}\gamma_{4}\right)
\end{equation}
Thus, Eq.~\eqref{rddot1} is approximated by a linear constant-coefficient ODE in terms of the small parameter $\xi(\tau)$, where $r_{0}\xi(t)$ represents the time-varying deviation from the orbit radius at epoch. 

Eq.~\eqref{xiODEp2} bears some structural resemblance to Eq.~\eqref{rddot1f} from the previous section. Both equations are linear oscillators with forcing terms due to the perturbations. In this case, Jacobi integral-dependent terms appear instead of energy, along with the addition of the $\gamma_{i}$ terms. This ODE is solved in the same way as the previous section, but the new $\overline{f}$ must first be defined succinctly in terms of $\tau$. Without loss of generality, the epoch time $t = 0$ can be defined at an instant when the first body principal axis aligns with the current line of nodes, $\hat{\bm{a}}_{1}\cdot\hat{\bm{\Omega}} = 1$. One may define $\hat{\bm{\gamma}}$ to point in this initial direction for all time, so $\Omega_{0} = 0$ by construction. Then $\psi = ct$ and $\theta \approx \theta_{0} + n_{0}t$ render $\overline{f}$ and $\delta i$ as explicit functions of time $t$. Other initializations are possible, but this is convenient because all initial system configurations may be captured by just two initial angles: $i_{0}$ and $\theta_{0}$. Note that the resulting equations using this convention may be simplified further for equatorial orbits, so for the special case of $i_{0} = 0$, there is no need for repeating the general derivation that follows. 

The substitution $t = \left(\mu/r_{0}^{3}\right)^{-1/2}\tau$ renders $\overline{f}$ and $\delta i$ as functions of the dimensionless time, where $\theta = \theta_{0} + \left(r_{0}^{3}/a_{0}^{3}\right)^{1/2}\tau$ and $\psi = c\left(\mu/r_{0}^{3}\right)^{-1/2}\tau$. The simplified expressions are given below, where $\alpha = C_{20}\left(R/r_{0}\right)^{2}$ and $\beta = C_{22}\left(R/r_{0}\right)^{2}$:
\begin{equation}
\label{f1oftau1}
\begin{split}
\overline{f}(\tau) = & \ \alpha\left[-\frac{1}{2} + \frac{3}{4}\sin^{2}{i_{0}}(1 - \cos{2\theta_{0}}\cos{\omega_{4}\tau} + \sin{2\theta_{0}}\sin{\omega_{4}\tau})\right] \\ & + 3\beta\bigg[\cos^{4}{\left(\frac{i_{0}}{2}\right)}\left(\cos{2\theta_{0}}\cos{\omega_{1}\tau} - \sin{2\theta_{0}}\sin{\omega_{1}\tau}\right) \\ & + \sin^{4}{\left(\frac{i_{0}}{2}\right)}\left(\cos{2\theta_{0}}\cos{\omega_{2}\tau} - \sin{2\theta_{0}}\sin{\omega_{2}\tau}\right) + \frac{1}{2}\sin^{2}{i_{0}}\cos{\omega_{3}\tau}\bigg]
\end{split}
\end{equation}
\begin{equation}
\label{deltaioftau1}
\begin{split}
\delta i(\tau) = & \ \frac{3}{8}\alpha\frac{\mu}{n_{0}h_{0}r_{0}}\bigg[\sin{2i_{0}}\left(\cos{2\theta_{0}}\left(1 - \cos{\omega_{4}\tau}\right) + \sin{2\theta_{0}}\sin{\omega_{4}\tau}\right)\bigg] \\ & + \frac{3}{4}\beta\frac{\mu}{ch_{0}r_{0}}\left(\frac{1}{c^{2}-n_{0}^{2}}\right)\bigg[-cn_{0}\sin{2i_{0}}\cos{2\theta_{0}} - 2\big(c^{2}\cos{2\theta_{0}} + (c^{2}-n_{0}^{2})\big)\sin{i_{0}} \\ & + \frac{1}{2}c(c+n_{0})\left(\sin{2i_{0}} + 2\sin{i_{0}}\right)\left(\cos{2\theta_{0}}\cos{\omega_{1}\tau} - \sin{2\theta_{0}}\sin{\omega_{1}\tau}\right) \\ & - \frac{1}{2}c(c-n_{0})\left(\sin{2i_{0}} - 2\sin{i_{0}}\right)\left(\cos{2\theta_{0}}\cos{\omega_{2}\tau} - \sin{2\theta_{0}}\sin{\omega_{2}\tau}\right) \\ & + 2(c^{2} - n_{0}^{2})\sin{i_{0}}\cos{\omega_{3}\tau}\bigg]
\end{split}
\end{equation}
\begin{equation}
\label{omegaP21}
\omega_{1} =  2(n_{0} - c)\sqrt{\frac{r_{0}^{3}}{\mu}}, \ \omega_{2} = 2(n_{0} + c)\sqrt{\frac{r_{0}^{3}}{\mu}}, \ \omega_{3} = 2c\sqrt{\frac{r_{0}^{3}}{\mu}}, \ \omega_{4} = 2n_{0}\sqrt{\frac{r_{0}^{3}}{\mu}} 
\end{equation}
Eq. \eqref{omegaP21} shows that the differential equation for $\xi$ is forced by four distinct frequencies if $n_{0} \neq c$. The solution to Eq.~\eqref{xiODEp2} is obtained in exactly the same way as in the previous section, with the final result given below, in terms of these four forcing frequencies and the oscillator natural frequency:
\begin{equation}
\label{xitauP2}
\xi(\tau) = \sum_{i = 1}^{4} A_{i}\cos{\omega_{i}\tau} + \sum_{i = 1, \ i\neq3}^{4}B_{i}\sin{\omega_{i}\tau} + C + D\cos{\sqrt{2\eta_{2}}\tau} + E\sin{\sqrt{2\eta_{2}}\tau}
\end{equation}
\begin{equation}
\label{A1f}
A_{1} = -3\beta\cos{2\theta_{0}}\left(\frac{\cos^{4}{\left(\frac{i_{0}}{2}\right)}\varphi - \frac{\mu}{8(c-n_{0})h_{0}r_{0}}\left(\sin{2i_{0}} + 2\sin{i_{0}}\right)\vartheta}{2\eta_{2} - \omega_{1}^{2}}\right)
\end{equation}
\begin{equation}
\label{A2f}
A_{2} = -3\beta\cos{2\theta_{0}}\left(\frac{\sin^{4}{\left(\frac{i_{0}}{2}\right)}\varphi + \frac{\mu}{8(c+n_{0})h_{0}r_{0}}\left(\sin{2i_{0}} - 2\sin{i_{0}}\right)\vartheta}{2\eta_{2} - \omega_{2}^{2}}\right)
\end{equation}
\begin{equation}
\label{A3f}
A_{3} = -\frac{3}{2}\beta\left(\frac{\sin^{2}{i_{0}}\varphi - \frac{\mu}{ch_{0}r_{0}}\sin{i_{0}}\vartheta}{2\eta_{2} - \omega_{3}^{2}}\right)
\end{equation}
\begin{equation}
\label{A4f}
A_{4} = \frac{3}{4}\alpha\cos{2\theta_{0}}\left(\frac{\sin^{2}{i_{0}}\varphi - \frac{\mu}{2n_{0}h_{0}r_{0}}\sin{2i_{0}}\vartheta}{2\eta_{2} - \omega_{4}^{2}}\right)
\end{equation}
\begin{equation}
\label{Bif}
B_{i} = -\left(\frac{\sin{2\theta_{0}}}{\cos{2\theta_{0}}}\right)A_{i}, \ i = 1,2,4
\end{equation}
\begin{equation}
\label{Cf}
C = \frac{\alpha}{4\eta_{2}}\left(1 - \frac{3}{2}\sin^{2}{i}\right)\varphi + \frac{\eta_{3}}{\eta_{2}} + \frac{3\alpha}{16}\left(\frac{\mu\sin{2i_{0}}\cos{2\theta_{0}}}{n_{0}h_{0}r_{0}\eta_{2}}\right)\vartheta
\end{equation}
\begin{equation}
\label{DEf}
D = -\sum_{i=1}^{4}A_{i} - C, \ E = \frac{1}{\sqrt{2\eta_{2}}}\left(\sqrt{\frac{r_{0}^{3}}{\mu}}\frac{\dot{r}_{0}}{r_{0}} - \sum_{i = 1, \ i\neq3}^{4}B_{i}\omega_{i}\right)
\end{equation}
Note the resonance condition $\omega_{h}^{2} = 2\eta_{2} = \omega_{i}^{2}$ captured by the denominators of $A_{i}$ and $B_{i}$ for $i = 1, 2, 3, 4$. Interestingly, these results also imply periodicity of $\xi(\tau)$, $\delta i (\tau)$, and $\overline{f}(\tau)$ if the resonance condition is avoided and $\omega_{i}/\omega_{j} \in \mathbb{Q} \ \forall i, j$ and $\omega_{i}/\omega_{h} \in \mathbb{Q} \ \forall i$, where $\mathbb{Q}$ denotes the set of rational numbers. 

The behavior of the orbit radius may be approximated as $r(\tau) = r_{0}(1 + \xi(\tau))$ using Eq.~\eqref{xitauP2} with the constants and frequencies defined above, and the transformation $\tau = \sqrt{\mu/r_{0}^{3}}t$ may be used to yield $r(t)$ explicitly. This and the previous constant-coefficient time-explicit expressions are analyzed and tested later in the paper with nonlinear truth model data.

\section{Expressions for Remaining Orbit Parameters}
With the inclination and orbit radius both approximated to $\mathcal{O}(\epsilon)$ by Eqs.~\eqref{iApprox_v1Solve} and \eqref{xitauP2} respectively, the approximations of variations in the other elements are now developed. Recall that in this paper, the orbit is parameterized by $\Omega, i, \theta, r, \omega_{n}, \dot{r}$. 

Some of the variations are direct results or analogs of the previous analysis. In particular, note that $\dot{r}$ is simply approximated by $r_{0}\dot{\xi}$, where $\dot{\xi}$ is given below:
\begin{equation}
\label{xi_dot_of_t}
\begin{split}
\dot{\xi}(\tau) = & \ \sqrt{\frac{\mu}{r_{0}^{3}}}\Bigg( -\sum_{i = 1}^{4} A_{i}\omega_{i}\sin{\omega_{i}\tau} + \sum_{i = 1, \ i\neq3}^{4}B_{i}\omega_{i}\cos{\omega_{i}\tau} - D\sqrt{2\eta_{2}}\sin{\sqrt{2\eta_{2}}\tau} \\
& + E\sqrt{2\eta_{2}}\cos{\sqrt{2\eta_{2}}\tau} \Bigg)
\end{split}
\end{equation}

The angular rate $\omega_{n}$ is already given in Eq.~\eqref{thetadotfromJ}, and can be explicitly obtained by substitution of the approximations for $i$, $r$, and $\dot{r}$ into Eq.~\eqref{thetadotfromJ}, while using $\theta \approx \theta_{0} + n_{0}t$ in $U_{2}(r)$. Only $\mathcal{O}(\epsilon)$ terms should be kept for consistency with the other approximate variations. The result is given below:
\begin{equation}
\label{thetadot_of_t}
\omega_{n} = c\cos{i_{0}} - \upsilon + \left(\frac{c^{2}\cos{i_{0}}\sin{i_{0}}}{\upsilon} - c\sin{i_{0}}\right)\delta i(t) + \frac{3\mu + 2J_{0}r_{0}}{r_{0}^{3}\upsilon}\xi(t) - \frac{\mu}{r_{0}^{3}\upsilon}\overline{f}(t)
\end{equation}
where $\upsilon$ is a function of initial conditions:
\begin{equation}
\label{upsilon}
\upsilon = \sqrt{\frac{2\left(\mu + J_{0}r_{0}\right)}{r_{0}^{3}} + c^{2}\cos^{2}{i_{0}}} 
\end{equation}

The variation in $\Omega$ is captured to $\mathcal{O}(\epsilon)$ in the same manner as the inclination:
\begin{equation}
\label{Omega_of_t}
\begin{split}
\Omega(t^{*}) \approx \Omega_{0} + \frac{3\mu R^{2}}{h_{0}r_{0}^{3}}\int_{0}^{t^{*}}\Big(C_{22}&\sin{(2(\Omega_{0}-\psi))}\sin{2\theta} \\ & + \left(C_{20} + 2C_{22}\cos{(2(\Omega_{0}-\psi))}\right)\cos{i_{0}}\sin^{2}{\theta}\Big){\D}t
\end{split}
\end{equation}
Reusing the free constraint $\Omega_{0} = 0$ from the approximation of the orbit radius, the equation for the variation in $\Omega$ is given below for near-circular orbits:
\begin{equation}
\label{Omega_of_t}
\begin{split}
\Omega(t) \approx & \ \frac{3\mu R^{2}}{h_{0}r_{0}^{3}}\Bigg(C_{20}\frac{\cos{i_{0}}}{2n_{0}}\left(\theta - \theta_{0} - \cos{\theta}\sin{\theta} + \cos{\theta_{0}}\sin{\theta_{0}}\right) \\ & - C_{22}\frac{1}{2c(c-n_{0})(c+n_{0})}\bigg(c\left(n_{0}\cos{2\theta}\sin{2\psi} - c\left(\cos{2\psi}\sin{2\theta} - \sin{2\theta_{0}}\right) \right)) \\ & + \cos{i_{0}}\left(\left(n_{0}^{2} - c^{2}(1 - \cos{2\theta})\right)\sin{2\psi} - cn_{0}\left(\cos{2\psi}\sin{2\theta} - \sin{2\theta_{0}}\right)\right)\bigg)\Bigg)
\end{split}
\end{equation}
This equation is evaluated by applying $\theta \approx \theta_{0} + n_{0}t$ and $\psi = ct$.

The argument of latitude $\theta$ is the final coordinate needed for parameterizing the orbit. Recall that the argument of latitude rate is given as
$\dot{\theta} = \omega_{n} - \dot{\Omega}\cos{i}$, where the second term is due to the deviation and regression of the node from which $\theta$ is measured \citep{PrussingConway2013}. The approximation for $\theta(t)$ is given by integrating the following equation, substituting Eq.~\eqref{thetadotfromJ} and the Gauss planetary equation for $\Omega$ and retaining only terms up to $\mathcal{O}(\epsilon)$:
\begin{equation}
\label{theta_of_t_int}
\theta(t^{*}) = \theta_{0} +  \int_{0}^{t^{*}}\left(\omega_{n}(t) - \dot{\Omega}(t)\cos{i_{0}}\right){\D}t
\end{equation}

Substituting preceding results into Eq. \eqref{theta_of_t_int} and simplifying:
\begin{equation}
\label{theta_of_t_int_v2}
\begin{split}
\theta(t^{*}) \approx & \ \theta_{0} + \left(c\cos{i_{0}} - \upsilon\right)t^{*} + \left(\frac{c^{2}\cos{i_{0}}\sin{i_{0}}}{\upsilon} - c\sin{i_{0}}\right)\int_{0}^{t^{*}}\delta i(t) \ {\D}t \\ & + \frac{3\mu + 2J_{0}r_{0}}{r_{0}^{3}\upsilon}\int_{0}^{t^{*}}\xi(t) \ {\D}t - \frac{\mu}{r_{0}^{3}\upsilon}\int_{0}^{t^{*}}\overline{f}(t) \ {\D}t - \Omega(t)\cos{i_{0}}
\end{split}
\end{equation}
where each integral expression is given below:
\begin{equation}
\label{integral_di}
\begin{split}
\int_{0}^{t^{*}}\delta i(t) \ {\D}t = & \ \sqrt{\frac{r_{0}^{3}}{\mu}}\Bigg(\ \frac{3}{8}\alpha\frac{\mu}{n_{0}h_{0}r_{0}}\bigg[\sin{2i_{0}}\left(\cos{2\theta_{0}}\left(\tau - \frac{\sin{\omega_{4}\tau}}{\omega_{4}}\right) - \sin{2\theta_{0}}\frac{\cos{\omega_{4}\tau}}{\omega_{4}}\right)\bigg] \\ & + \frac{3}{4}\beta\frac{\mu}{ch_{0}r_{0}\left(c^{2}-n_{0}^{2}\right)}\bigg[-cn_{0}\sin{2i_{0}}\cos{2\theta_{0}}\tau - 2\big(c^{2}\cos{2\theta_{0}} + (c^{2}-n_{0}^{2})\big)\sin{i_{0}}\tau \\ & + \frac{1}{2}c(c+n_{0})\left(\sin{2i_{0}} + 2\sin{i_{0}}\right)\left(\cos{2\theta_{0}}\frac{\sin{\omega_{1}\tau}}{\omega_{1}} + \sin{2\theta_{0}}\frac{\cos{\omega_{1}\tau}}{\omega_{1}}\right) \\ & - \frac{1}{2}c(c-n_{0})\left(\sin{2i_{0}} - 2\sin{i_{0}}\right)\left(\cos{2\theta_{0}}\frac{\sin{\omega_{2}\tau}}{\omega_{2}} + \sin{2\theta_{0}}\frac{\cos{\omega_{2}\tau}}{\omega_{2}}\right) \\ & + 2(c^{2} - n_{0}^{2})\sin{i_{0}}\frac{\sin{\omega_{3}\tau}}{\omega_{3}}\bigg]\Bigg)
\end{split}
\end{equation} 
\begin{equation}
\label{integral_xi}
\begin{split}
\int_{0}^{t^{*}}\xi(t) \ {\D}t = & \ \sqrt{\frac{r_{0}^{3}}{\mu}}\Bigg( \sum_{i = 1}^{4} \frac{A_{i}}{\omega_{i}}\sin{\omega_{i}\tau} - \sum_{i = 1, \ i\neq3}^{4}\frac{B_{i}}{\omega_{i}}\cos{\omega_{i}\tau} + C\tau + \frac{D}{\sqrt{2\eta_{2}}}\sin{\sqrt{2\eta_{2}}\tau} \\ & - \frac{E}{\sqrt{2\eta_{2}}}\cos{\sqrt{2\eta_{2}}\tau}\Bigg)
\end{split}
\end{equation}
\begin{equation}
\label{integral_f1}
\begin{split}
\int_{0}^{t^{*}}\overline{f}(t) \ {\D}t = & \ \sqrt{\frac{r_{0}^{3}}{\mu}}\Bigg( \alpha\left[-\frac{1}{2}\tau + \frac{3}{4}\sin^{2}{i_{0}}(1 - \cos{2\theta_{0}}\frac{\sin{\omega_{4}\tau}}{\omega_{4}} - \sin{2\theta_{0}}\frac{\cos{\omega_{4}\tau})}{\omega_{4}}\right] \\ & + 3\beta\bigg[\cos^{4}{\left(\frac{i_{0}}{2}\right)}\left(\cos{2\theta_{0}}\frac{\sin{\omega_{1}\tau}}{\omega_{1}} + \sin{2\theta_{0}}\frac{\cos{\omega_{1}\tau}}{\omega_{1}}\right) \\ & + \sin^{4}{\left(\frac{i_{0}}{2}\right)}\left(\cos{2\theta_{0}}\frac{\sin{\omega_{2}\tau}}{\omega_{2}} + \sin{2\theta_{0}}\frac{\cos{\omega_{2}\tau}}{\omega_{2}}\right) + \frac{1}{2}\sin^{2}{i_{0}}\frac{\sin{\omega_{3}\tau}}{\omega_{3}}\bigg]\Bigg)
\end{split}
\end{equation}

At this point, the approximate behaviors of all 6 state elements $\Omega, i, \theta, r, \omega_{n}, \dot{r}$ are fully developed. The necessary information for the first 5 elements are given respectively in Eqs.~\eqref{Omega_of_t}, \eqref{deltaioftau1}, \eqref{theta_of_t_int_v2}, \eqref{xitauP2}, and \eqref{thetadot_of_t}. To use these equations, the reader is reminded of the definitions $i(\tau) = i_{0} + \delta i(\tau)$, $r(\tau) = r_{0}(1 + \xi(\tau))$, and $\dot{r} = r_{0}\dot{\xi}(\tau) = r_{0}\sqrt{\mu/r_{0}^{3}}\xi'(\tau)$, where $\tau$ is given by Eq.~\eqref{Ttransform1}. Differentiation of Eq.~\eqref{xitauP2} is straightforward so $\xi'(\tau)$ is not explicitly given.  These results enable near-circular orbits in the rotating gravity field to be analytically approximated. The elements $\Omega, i, \theta, r$ capture the position, and the elements $\dot{r}$ and $\omega_{n}$ capture the velocity. The approximations for $\Omega$ and $i$ are easy to obtain, so they might exist elsewhere in literature. To the knowledge of the authors, the other expressions in this paper appear here for the first time. All elements are tested numerically in this paper, but most of the focus is on studying the accuracy of the approximation of the orbit radius, whose accuracy will generally reflect the accuracy of approximations of $\theta$ and $\omega_{n}$ due to the coupling of these quantities. 

\subsection{Periodicity of the Perturbed Elements}
Recall the periodicity condition on $\xi$, $\delta i$, and $\overline{f}$ is given as any admissible choice of angular frequencies $\omega_{h}$, $\omega_{i}$ such that the resonance condition $\omega_{h}^{2} = 2\eta_{2} = \omega_{i}^{2}$ is avoided and $\omega_{i}/\omega_{j} \in \mathbb{Q} \ \forall i, j$ and $\omega_{i}/\omega_{h} \in \mathbb{Q} \ \forall i$. When this condition is satisfied, the analytic solutions predict that the elements $r$, $\dot{r}$, $\omega_{n}$, and $i$ will be periodic. The variation of $\Omega$ has the following secular rate $\dot{\overline{\Omega}}$:
\begin{equation}
\label{Omega_sec}
\dot{\overline{\Omega}} = \frac{3\mu R^{2}}{2n_{0}h_{0}r_{0}^{3}}C_{20}\cos{i_{0}}
\end{equation}
Common periodicity of the remaining elements $\theta$ and $\Omega$ could only be achieved by a choice of initial conditions satisfying the periodicity condition of $\xi$, $\delta i$, and $\overline{f}$, resulting in a common period $T^{*}$ for which the elements $r$, $\dot{r}$, $\omega_{n}$, and $i$ are periodic, with the additional constraint that this $T^{*}$ must satisfy $\theta(t_{0} + T^{*}) = \theta(t_{0}) + 2\pi k$. In addition, the secular right ascension drift $\Delta\Omega$ over $k$ orbits must be accounted for, i.e. $2\pi k = \psi + |\Delta\Omega|$, resulting in the following final constraint for bodies with $C_{20} < 0$:
\begin{equation}
\label{Omega_const}
kn_{0} - \left(c - \frac{3\mu R^{2}}{2n_{0}h_{0}r_{0}^{3}}C_{20}\cos{i_{0}}\right)\left(k - \frac{1}{l}\right) = 0
\end{equation}
where the perturbed element common period is represented as $T^{*} = \left(k - \frac{1}{l}\right)T_{0}$ for integers $k$ and $l$ satisfying $k \geq 1$, $|l| > 1$. Inspection of Eq. \eqref{Omega_const} implies that for $\Gamma > 1$ and $C_{20} < 0$, the negative drift in $\Omega$ cannot be accounted for by prograde orbits, and conditions for full periodicity of the elements cannot be found. However, using the analytic results in this paper, prograde orbits that are periodic except for their precession can be found with relative ease. Such orbits could have useful applications for asteroid mission planning.

\section{Limitations and Extensions}
The derivation in this paper assumes the orbit is near-circular, and that terms due to the potential (captured by $\overline{f}(t)$) and the deviations $\delta i$ and $\xi(t)$ manifest at the same order in the dimensionless equations. This is not always the case, and the current treatment is inappropriate in particular for very high inclination orbits. Furthermore, the method for approximating the solution to Eq.~\eqref{rddot1} assumes that the variations in the perturbed orbit radius remain relatively small. This assumption results in poor solution accuracy as the critical value $\Gamma = 1$ is approached, as well as the appearance of singularities in the coefficients for $\xi(\tau)$ for when $\Gamma$ is unity. Additionally, Eq.~\eqref{eta2} breaks down somewhere before $\Gamma = 1$, when $\eta_{2}$ becomes imaginary, so the frequency $\sqrt{2\eta_{2}}$ of the homogeneous solution to Eq.~\eqref{xiODEp2} also fails. Examination of Eq.~\eqref{eta2} reveals the critical condition $\gamma_{1} + 2\gamma_{2} = 0$, which can be approximated as $\Gamma^{2}\cos^{2}{i} - 2\Gamma\cos{i} \approx - 1$. Thus, failure occurs near when $\Gamma = 1/\cos{i}$. 

For similar reasons, high inclination can drive solution failure at higher levels of $\Gamma$. 
By investigating the offending term $\eta_{2}$, one can obtain the following approximate inclination limit for accurately modeling prograde orbits: 
\begin{equation}
\label{g1_crit}
i_{0} \lesssim \cos^{-1}{\left(\frac{1}{\Gamma}\right)}
\end{equation}
Solution accuracy will start to degrade as the inclination approaches this value from below. For high values of $\Gamma$, this inclination limit is relatively high. For example, when $\Gamma = 4$, the solution is valid for inclinations in the range $|i| < 75^{\circ}$.

Because this work only examined linearized oscillations in the perturbed orbit, a logical extension would be to consider the case of nonlinear oscillations. 
The solutions in this paper neglect the nonlinearities in $\xi$, which become important for larger values of $\xi$. For example, Eq. \eqref{rddotP21} could be expanded to quadratic order in $\xi$.
There are well-established techniques for examining weakly nonlinear oscillators, discussed extensively in \cite{nayfeh2008nonlinear}.

The work in this paper is focused on approximating perturbed orbits that are nearly circular as $r \approx r_{0}(1 + \xi(t))$, where time is used as the independent variable to take advantage of the near-constant angular rate $\omega_{n}$. An attractive analog of this approach for the eccentric orbit problem would be to let $\xi$ accommodate the deviations in the perturbed eccentric radius from the unperturbed case:
\begin{equation}
\label{r_theta_v2}
r(\theta) = \frac{a_{0}(1-e_{0}^{2})}{1 + e_{0}\cos{\left(\theta - \omega_{0}\right)}}\left(1 + \xi(\theta)\right)
\end{equation}
For this analysis, the independent variable of Eq.~\eqref{rddot1} is transformed from $t$ to $\theta$ through the following substitution:
\begin{equation}
\label{dtheta1}
\frac{{\D}}{{\D} t}\left( \ \right) = \dot{\theta}\frac{{\D}}{{\D}\theta}\left( \ \right) = \left(\omega_{n} - \dot{\Omega}\cos{i}\right)\frac{{\D}}{{\D}\theta}\left( \ \right)
\end{equation}
For this eccentric case, one would expect that the only feasible orbit approximations in terms of time $t$ (instead of an anomaly) would be lengthy and inconvenient series expressions, similar in spirit to expansions that can be found for the Keplerian problem in \cite{battin1}. By switching the independent variable to $\theta$, far fewer frequencies are needed to approximate the solution to a desired level of accuracy, resulting in simpler solutions. However, the transformation given by Eq. \eqref{dtheta1} introduces the additional inconvenience of having to express the time dependence for the angle $\psi = \psi_{0} + ct$ as a function of $\theta$. Additionally, for a solution $\xi(\theta)$ to be valid for a range of eccentricities $0 \leq e_{0} < 1$, it must perform different characteristic roles depending on the scale of $e_{0}$. For $e_{0} = 0$, it should behave like the original solution in this paper. For low values of $e_{0}$, the perturbed argument of periapsis oscillates quickly through a wide range of angles \citep{Izsak1961}, so the leading term in Eq. \eqref{r_theta_v2} is inaccurate and must essentially be compensated for by $\xi(\theta)$. For higher values of $e_{0}$, the variations in the argument of periapsis are greatly reduced, and the variations of $r$ with $\theta$ will behave more like the leading term in Eq. \eqref{r_theta_v2}. In this final case, an admissible universal solution $\xi(\theta)$ should account for the small perturbation-induced deviations in orbit radius.


\section{Applications}
Even with the limitations discussed in Section 4, this new model is applicable to a large number of cases in nature. For asteroids, the $C_{20}$ and $C_{22}$ coefficients are often dynamically important \citep{Scheeres}, and can be the dominant disturbances if the asteroid is large enough for its gravity to overpower solar radiation pressure and solar gravity disturbances for close orbits. Past studies have demonstrated that if the semimajor axis is above a 1.5 resonance radii limit and below a corresponding upper limit characterized by solar radiation pressure (SRP) perturbations, the orbit will be more likely to persist for long time spans \citep{HuScheeres2004,ScheeresJGCD2012,ScheeresMarzari2002}:
\begin{equation}
\label{SMA_limits}
\frac{3}{2}\left(\frac{T_{r}^{2}\mu}{4\pi^{2}}\right)^{1/3} < a < \frac{1}{4}\sqrt{\frac{\mu B}{G_{1}}}d
\end{equation}
Here $T_{r}$ is the rotation period of the asteroid, $\mu$ is its gravitational parameter, $B$ is the spacecraft mass-to-area ratio in kg/m\textsuperscript{2}, $G_{1} \approx 10^{8} \text{kg}\cdot\text{km}^{3}/\text{s}^{2}\text{m}^{2}$ is the solar constant, and $d$ is the distance from the asteroid to the sun in km. Note also that the semimajor axis can be expressed as a function of the parameter $\Gamma = c/n_{0}$:
\begin{equation}
\label{a_from_Gamma}
a = \Gamma^{2/3}\left(\frac{T_{r}^{2}\mu}{4\pi^{2}}\right)^{1/3}
\end{equation}
Thus the inequality to guard against the effects of the rotating gravity field is $\Gamma > \left(3/2\right)^{3/2} \approx 1.8$. A fully analytic model for the $C_{20}$ and $C_{22}$ problem developed for $\Gamma$ significantly below this value would have limited applicability due to the influences of other gravitational harmonics. Also, the complicated dynamics in that region are not well-suited to approximation by modeling a small deviation from nominal behavior.

\begin{figure}[h!]
    \centering
    \includegraphics[]{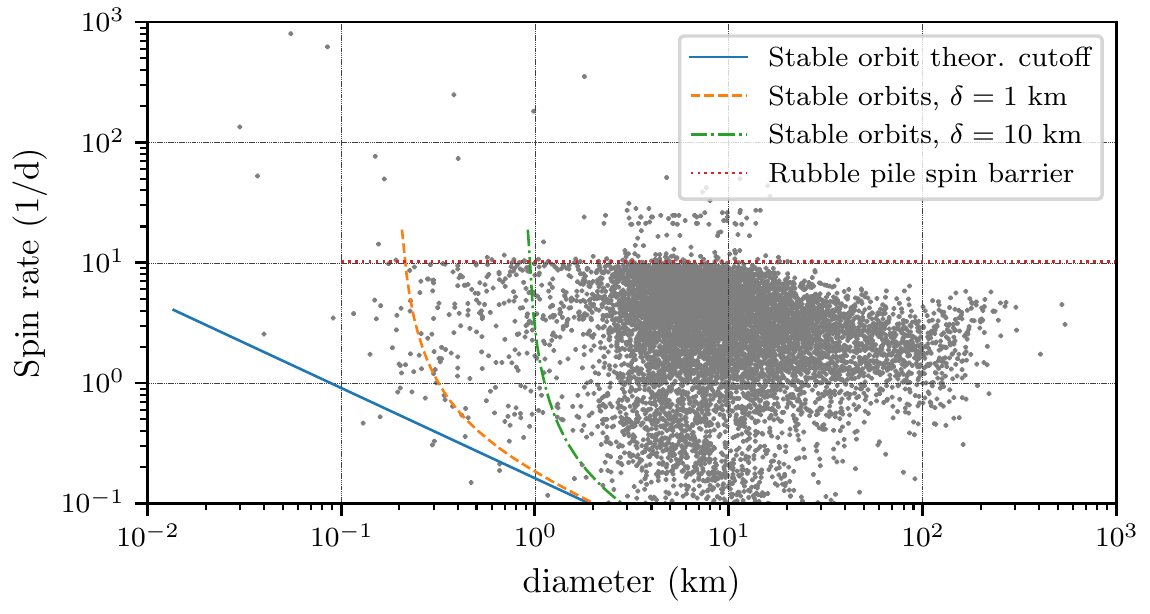}
    \caption{Orbital stability cutoffs for a selection of catalogued asteroids. This figure includes theoretical cutoffs for orbital stability for a large sample of asteroids from the Main Belt and Atira, Aten, Apollo, and Amor families.}
    \label{fig:ManyAsteroids}
\end{figure}

It can be shown that in the solar system's vast population of asteroids, a large number of them have regions above their surfaces that can fit orbits satisfying the stability condition given by Eq. \eqref{SMA_limits} -- that is, regions that are well-approximated by the new model in this paper. To begin, consider the limiting case that the upper and lower limits in Eq. \eqref{SMA_limits} are equal. In that case, the critical value of the asteroid gravitational parameter is given as a function of body rotation period:
\begin{equation}
\label{mu_crit}
\mu^{*} = \frac{6^{6}}{d^{6}}\frac{G_{1}^{3}}{B^{3}}\left(\frac{T_{r}^{2}}{4\pi^{2}}\right)^{2}
\end{equation}
The mean diameter of the asteroid $D$ is given as a function of $\mu$, the body density $\rho$, and the gravitational constant $G$:
\begin{equation}
\label{D_crit}
D^{*} = 2\left(\frac{3\mu^{*}}{4\pi\rho G}\right)^{1/3}
\end{equation}
In a figure of asteroid spin rates vs. diameters, Eq. \eqref{D_crit} gives a curve of the lower theoretical limit for an asteroid to have a stable region with orbits above 1.5 times the resonance radius and below the SRP disturbance limit. This curve is given in blue in Figure \ref{fig:ManyAsteroids}, along with spin-diameter data for a sample of 9400 asteroids obtained from the JPL small body database, mainly from the asteroid belt. This is done by assuming $\rho \sim 2$ g/cm\textsuperscript{3}, specifying $d = 2.7$ AU, and choosing $B = 20$ kg/m\textsuperscript{2}, which is a representative mass-to-area ratio for spacecraft that have visited asteroids \citep{BroschartEtAl2014}. However, orbital stability is much more likely if there exists a significant nonzero $\delta = r_{\text{max}} - r_{\text{min}}$, where $r_{\text{max}}$ and $r_{\text{min}}$ are the upper and lower limits in Eq. \eqref{SMA_limits}. Cutoff curves are given in Figure \ref{fig:ManyAsteroids} for $\delta = 1$ km and for $\delta = 10$ km. From these curves, it is seen that most asteroids with diameters above 1 km have at least 10 km of altitude difference between their upper and lower orbital stability limits. The region to the right of the cutoff curves corresponds to environments with stable orbital environments above 1.5 resonance radii ($\Gamma = 1.8$), so it is clear that the newly developed analytical solution (developed for $\Gamma > 1$) could find application at a large number of solar system bodies, including many of the main belt asteroids. 

Because the model does not consider the solar gravitational or radiation pressure disturbances, it will be most applicable for satellite orbits with $a \ll r_{max}$ to minimize unmodeled SRP effects, and $a \ll r_{H}$ to minimize unmodeled solar gravitational disturbances, where $r_{H}$ is the Hill radius:
\begin{equation}
\label{r_Hill}
r_{H} = d\left(\frac{\mu}{3\mu_{s}}\right)^{1/3}
\end{equation}
and $\mu_{s}$ is the solar gravitational parameter. The minimum diameter to achieve a given Hill radius is given below:
\begin{equation}
\label{D_from_Hill}
D > \left(\frac{18\mu_{s}}{\pi\rho G}\right)^{1/3}\left(\frac{r_{H}}{d}\right)
\end{equation}
Using this equation, it is shown that a Hill radius of 150 km can be achieved by main belt asteroids with $\rho = 2.0$ g/cm\textsuperscript{3} and $D > 663$ m. 
Note lastly that the ratio $r_{H}/D$ depends on $\rho$ and $d$, but not on $D$. It turns out that except for very dull spacecraft, the solar radiation pressure will often be more important than the solar gravitational disturbance.
Overall, this analysis shows that there are many possibilities to design a near-circular orbit above the lower stability limit, below the SRP limit, and well below the Hill sphere radius. If these conditions are satisfied, and if the inclination is below the polar limits, the motion will be governed by dynamics that are well-approximated by this new model. 

The model can be applied to the case of orbiters in near-circular orbits about sufficiently large asteroids. It can additionally be applied to the case of small moons in binary asteroid systems (i.e., in cases where $m \ll M$). Many times, these orbits are near-circular. Consider an object like the trinary asteroid (136617) 1994 CC. The primary body has a diameter of 650-700 m, and the $\sim 50$ m moon ``Beta" orbits with an eccentricity of just $e \approx 0.002$. With the primary body rotating with $T_{r} = 2.39$ hrs and the moon orbiting with a period of 1.24 days, the rotation rate ratio is $\Gamma \approx 12.5$. The orbit of this tiny moon would also be well-approximated by the model developed here. 
There are many other possible examples, because binary asteroids are quite common in the solar system, constituting up to 15\% of  near-Earth asteroids over 200 m.

\section{Numerical Simulations}
\subsection{Validating The Orbit Approximations}
Simulations confirm that the approximations of the orbit work as expected, and this section presents representative examples to demonstrate this. For all cases in this section, the hypothetical asteroid described in Table~\ref{table:Sim1Data} is used. It is a fairly representative example of main belt asteroids.

\begin{table}[h!]
\centering
\caption{Primary Body Parameters}
\label{table:Sim1Data}
\begin{tabular}{l|c}
Parameter                                                                                      & Value                                                                                          \\ \hline
Size and Mass Data & $R = 6.0 \ \text{km}, \ \rho = 2.6 \ \text{g}/\text{cm}^{3}, \ m = 4.9009\times10^{14} \ \text{kg}$ \\
Gravity Field Data & $ \mu = 3.2709\times10^{-5} \ \text{km}^{3}\text{s}^{-2}, \ C_{00} = 1.0, \ C_{20} = -0.0903, \ C_{22} = 0.0375$ \\
Rotation data & $10.0 < T_{r} < 50.0$ hours, constant variable. $\psi_{0} = 0.0$
\end{tabular}
\end{table}

\begin{figure}[h!]
	\centering
	\includegraphics[]{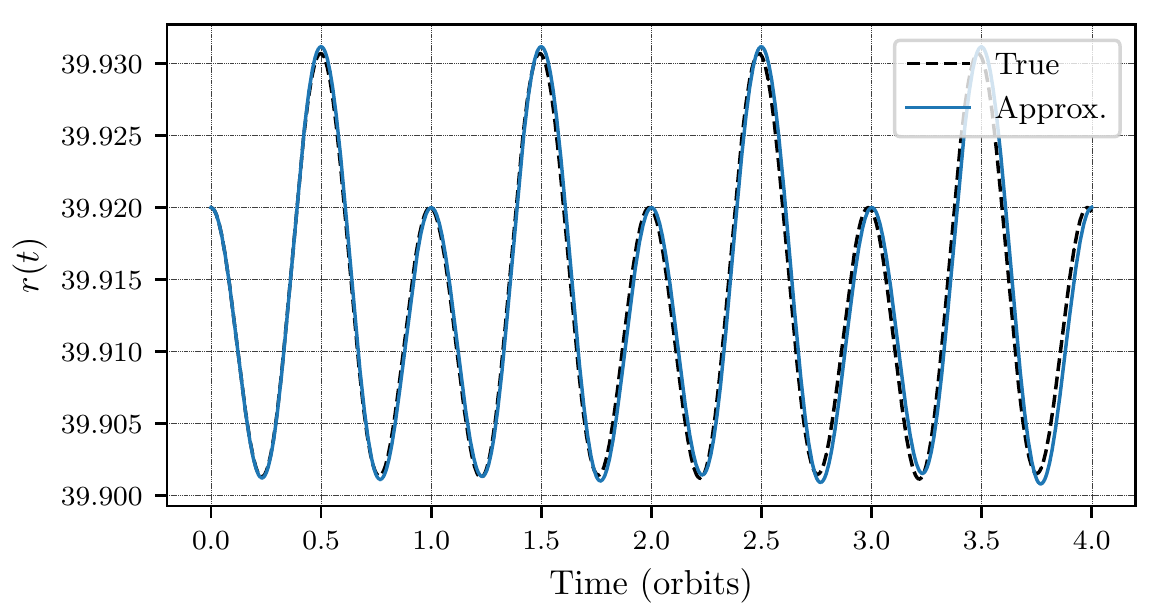}
	\caption{Orbit radius vs. time for Simulation 1, $C_{20}$ only. Time is in unperturbed orbit periods. The variations in $r$ are induced by nonzero initial eccentricity and by the gravitational perturbation from oblateness. For the $C_{20}$-only problem, only equatorial orbits can be purely circular.}
	\label{fig:r_of_t_P1a}
\end{figure}

For the first simulation, $C_{22} = 0$, and the $J_{2}$-only approximation of the orbit radius is tested. The initial nonzero orbital elements are $a_{0} = 40 \ \text{km}$, $e_{0} = 0.002$, $i_{0} = 50.0^{\circ}$. The resulting approximation of the orbit radius is compared to truth model data in Figure~\ref{fig:r_of_t_P1a}.  The results show that the approximation obtained with energy conservation works as expected with small initial eccentricity. 
\begin{figure}[h!]
	\centering
	\includegraphics[]{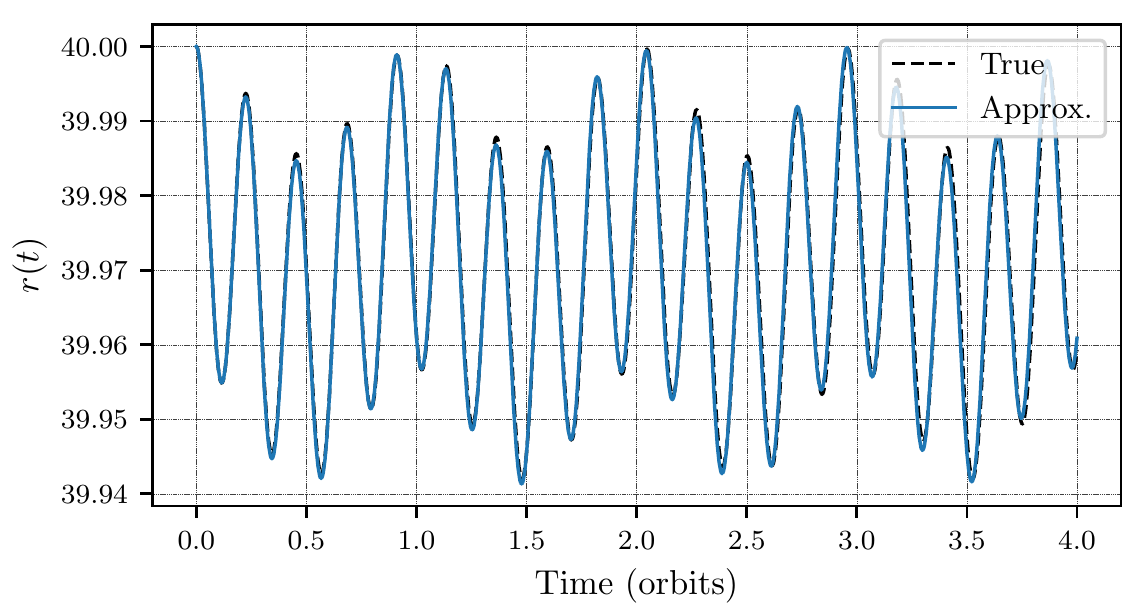} 
	\caption{Orbit radius vs. time for Simulation 2, $C_{20} + C_{22}, \ \Gamma = 3.2$, with time in unperturbed orbit periods. With this high value of $\Gamma$, there are small rapid variations in $r$ induced by the rotating gravity field. These are well-captured by the analytic model.}
	\label{fig:r_of_t_P2a}
\end{figure}
\begin{figure}[h!]
	\centering
	\includegraphics[]{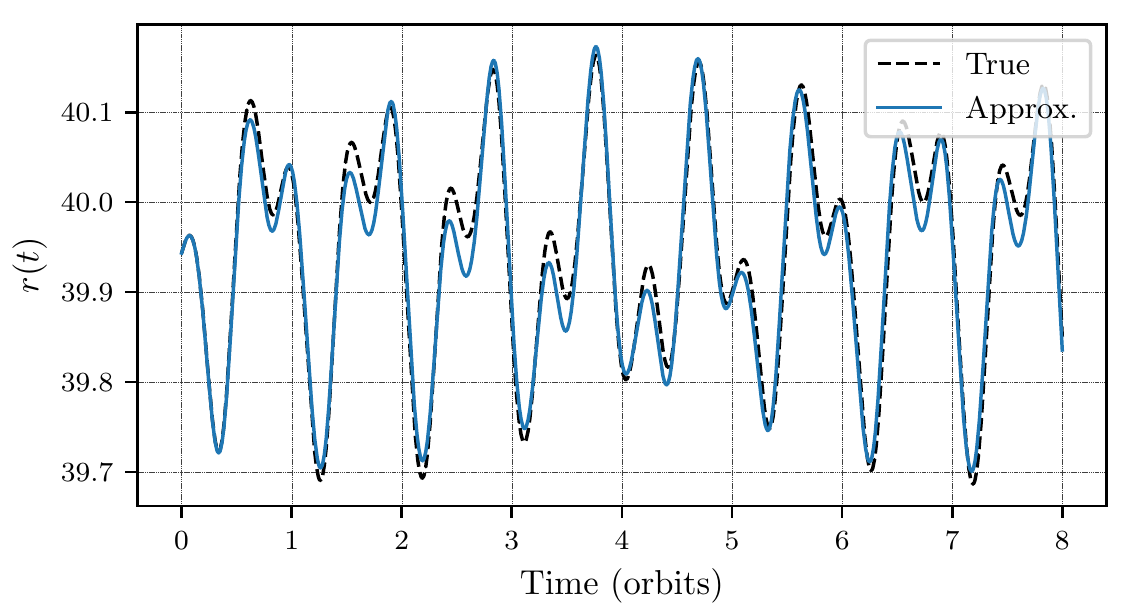} 
	\caption{Orbit Radius vs. time for Simulation 3, $C_{20} + C_{22}, \ \Gamma = 2.1$, with time in unperturbed orbit periods. With this lower value of $\Gamma$, the perturbation-induced variations in $r$ are larger, but are still characteristically well-approximated by the analytic model.}
	\label{fig:r_of_t_P2b}
\end{figure}
Simulation 2 uses $a_{0} = 40$ km, $e_{0} = 0$, and $i_{0} = 2.0^{\circ}, \theta_{0} = 0^{\circ}$. Simulation 3 uses $a_{0} = 40$ km, $e_{0} = 0.0022$, $i_{0} = 40.0^{\circ}, f_{0} = \theta_{0} = 50.0^{\circ}$. In all cases, the unperturbed orbit period is 77.2 hours. In the first case, the asteroid rotation period is set to 24.12 hours, resulting in the angular rate ratio $\Gamma = c/n = 3.2$. For this first case, the approximation of the orbit radius is compared to truth model data in Figure~\ref{fig:r_of_t_P2a}. In the second case, the asteroid rotation period is set to 36.76 hours, so $\Gamma = 2.1$. The results are given in Figure~\ref{fig:r_of_t_P2b}. 

In general, for lower values of $\Gamma$, the effects on variations in the orbit radius are more severe. In these cases of near-resonance, the fluctuations can be so large as to result in orbit ejection or impact with the primary body in the long-term. Such cases are not well-represented by any approximation assuming small deviations from the initial orbit radius. The third simulation with $\Gamma = 2.1$ is near the limit of efficacy of the current approach at this time of writing. The fluctuations are larger than in other cases, and the approximation is less accurate, while still predicting the general behavior. Overall, this approximation accuracy is limited to inclinations below $70^{\circ}$, as discussed earlier.
\begin{figure}[h!]
	\centering
	\includegraphics[]{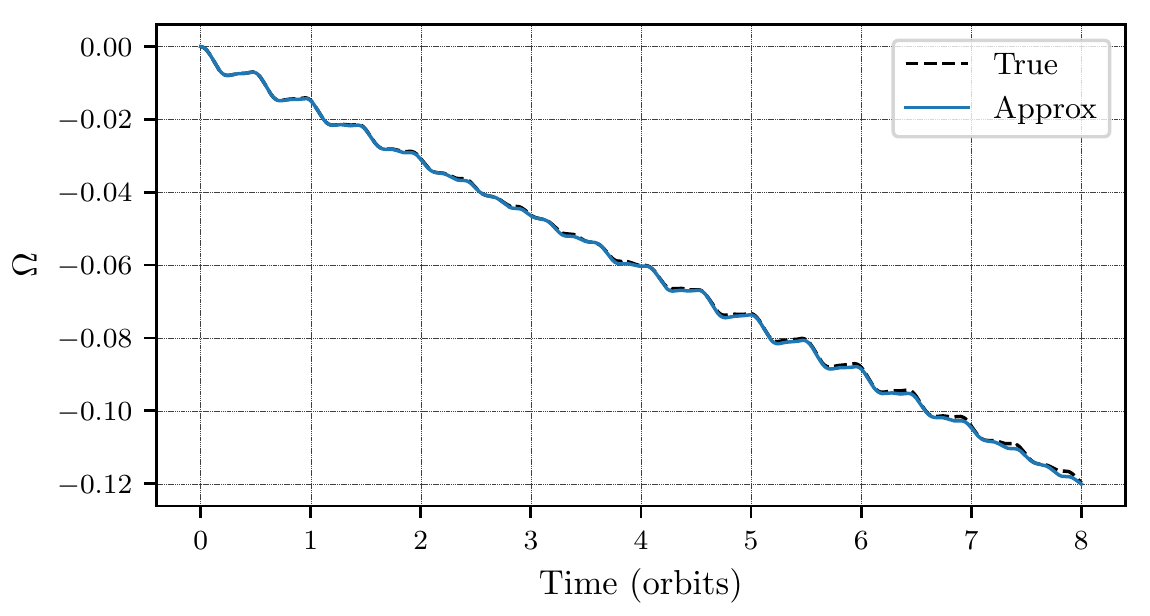} 
	\caption{Right Ascension vs. time for Simulation 3, $C_{20} + C_{22}, \ \Gamma = 2.1$. Using Eq. \eqref{Omega_of_t}, the secular drift in $\Omega$ and the short-period oscillations are well-captured.}
	\label{fig:raan_of_t_P2b}
\end{figure}
\begin{figure}[h!]
	\centering
	\includegraphics[]{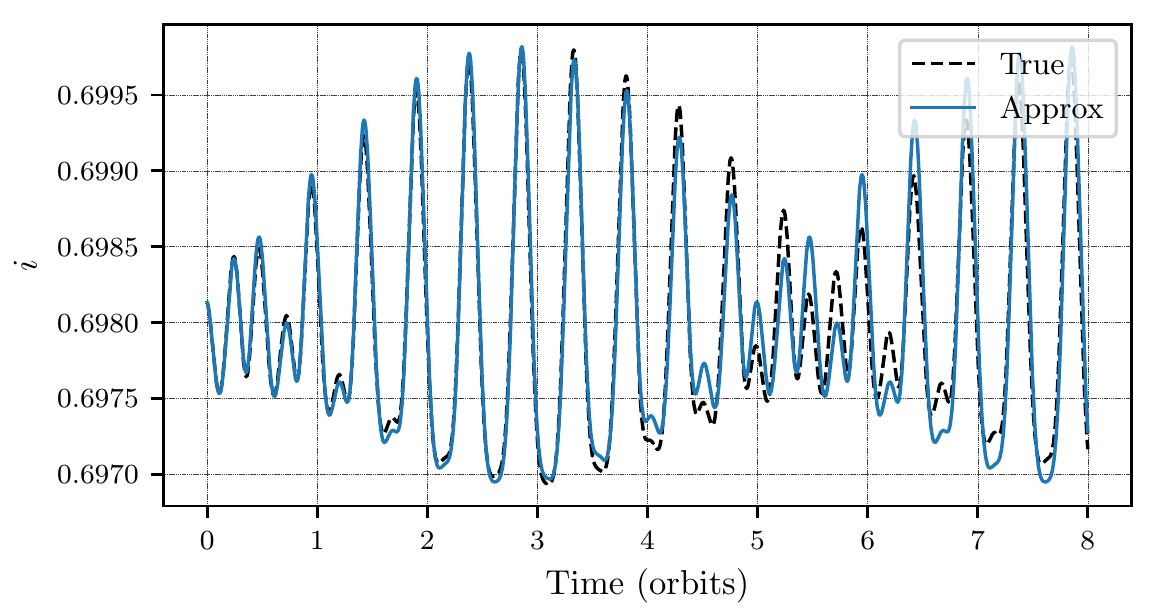} 
	\caption{Inclination vs. time for Simulation 3, $C_{20} + C_{22}, \ \Gamma = 2.1$. Using Eq. \eqref{iApprox_v1Solve}, the beating phenomenon in inclination variation is well-approximated over 8 unperturbed orbit periods.}
	\label{fig:inc_of_t_P2b}
\end{figure}
\begin{figure}[h!]
	\centering
	\includegraphics[]{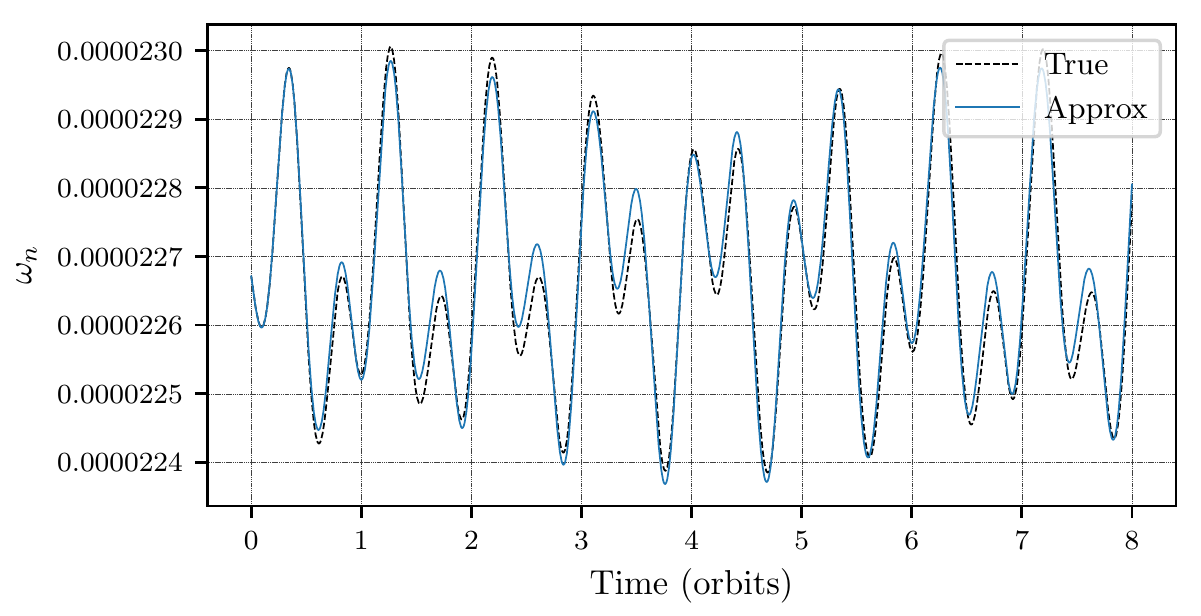} 
	\caption{Orbit-normal angular velocity vs. time for Simulation 3, $C_{20} + C_{22}, \ \Gamma = 2.1$. The accuracy in the angular rate approximation is comparable to the accuracy of the approximation of the radius given by Figure \ref{fig:r_of_t_P2b}.}
	\label{fig:thetadot_of_t_P2b}
\end{figure}

\begin{figure}[h!]
	\centering
	\includegraphics[]{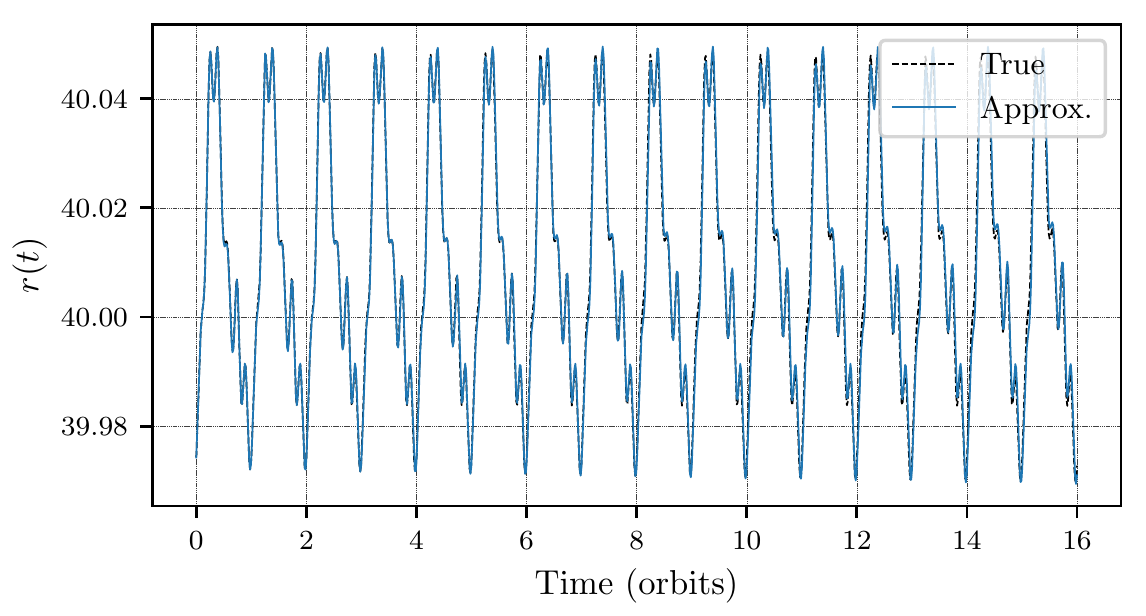} 
	\caption{Orbit Radius vs. time for Simulation 4, $C_{20} + C_{22}, \ \Gamma = 4.0$. This result shows that the analytic model is able to capture the small variations in orbit radius for 16 unperturbed orbit periods to a high degree of accuracy. This result generally holds for large values of $\Gamma$.}
	\label{fig:r_of_t_P3b}
\end{figure}

With the accuracy of the approximation of the orbit radius demonstrated for several examples, some representative results are shown to demonstrate that the other analytic approximations work. Figures \ref{fig:raan_of_t_P2b} -- \ref{fig:thetadot_of_t_P2b} show that the approximations of the right ascension, inclination, and $\omega_{n}$ are reasonably accurate for the case of Simulation 3, with $a_{0} = 40$ km, $e_{0} = 0.0022$, $i_{0} = 40.0^{\circ}, \theta_{0} = 50.0^{\circ}$. These are the same initial conditions as are used to generate Figure \ref{fig:r_of_t_P2b}. It would be somewhat redundant to show simulation results of $\dot{r}$ and $\theta$, so these results are omitted. 

One final simulation demonstrates the efficacy of the approximation for long time spans for cases with highly regular motion. In this simulation, $a_{0} = 40$ km, $e_{0} = 0.001$, $i_{0} = 50.0^{\circ}, \omega_{0} = 25.0^{\circ}, f_{0} = 50.0^{\circ}$. Furthermore, $\Gamma = 4.0$. The resulting motion is simulated for 16 orbits with the nonlinear dynamics and with the approximation, and the results agree to high accuracy for the full timespan simulated. Only the radius data is shown for brevity, and these results are given in Figure \ref{fig:r_of_t_P3b}. Note that the approximation captures the interesting feature of long-term variations in the brief sharp oscillations appearing $3/4$ of the way through each orbit, as well as the persistent larger orbit-periodic variations due to the initial nonzero eccentricity.

\section{Conclusions}
In this paper, the variations in the orbit radius in a rotating gravity field with $C_{20}$ and $C_{22}$ are described for near-circular orbits with the angular rate ratio $\Gamma = c/n > 1$. The scalar differential equation for the orbit radius $r$ is rendered as a time-varying differential equation in $r$ alone using the Jacobi integral to remove unknown terms to first order in small variational terms, $\mathcal{O}(\epsilon)$. Once the approximation for the orbit radius is obtained, approximations for all other components of the orbit state are found. The approximations in this paper are all explicit functions of time. Most time-dependent terms are weighted sums of $\sin{( \ )}$ and $\cos{( \ )}$, with the weights determined by system initial conditions, and 5 fundamental frequencies constructed from the initial mean motion $n_{0}$ and the primary body angular rotation rate $c$. Solution accuracy generally increases as $\Gamma$ is increased further above the critical value $\Gamma = 1$, with generally good accuracy especially for $\Gamma > 2$. This makes these solutions especially well-suited for approximating near-circular orbits around quickly rotating bodies with significant $C_{20}$ and $C_{22}$ coefficients. An analysis shows that this model applies to the dynamics in the vicinity of many asteroids, particularly those on the km scale.

The potential for additional analytical work is extensive. 
First, it would be useful to obtain an equivalent solution to the one derived in this work for retrograde instead of prograde orbits. 
The relative scale of variations in the derivation could be treated more formally. This would help in identifying which issues in the approximate solution are fundamental to this approach and which can be amended. Furthermore, a rigorous accounting of the relative scale of small terms would enable higher-order analytic series approximations of the orbit behavior to be obtained. The approximation of variations in the orbit radius can be extended to more eccentric orbits by a change of independent variable in Eq.~\eqref{rddot1}. The assumption that $|\omega_{n}| \gg |\dot{\Omega}|, |\dot{i}|$ should also be relaxed for highly eccentric orbits. 

For the problem of finding approximate solutions to the orbit problem perturbed by oblateness and ellipticity, any given approximate solution will only apply to limited regions of the parameter space of all possible orbits. For example, the solutions in this paper are applicable only to orbits that remain near-circular. The assumptions used to generate an approximation necessarily constrain its applicability to the space in which these assumptions are valid. The parametric variation of behavior of orbital motion in rotating asymmetric gravity fields has been extensively studied numerically \citep{HuScheeresCJA, Scheeres, Scheeres4769Castalia1996}. Numerical studies reveal that depending on the value of the parameter $\Gamma = c/n$, the orbits in the $C_{20}$ and $C_{22}$-perturbed problem can behave quite differently, with highly irregular behavior possible, including resonances. For such cases, analytic approximations assuming small deviations are certain to fail. However, analytic approximation techniques could be successful in revealing the boundaries of such irregular regions in the parameter space. 

\bibliographystyle{cas-model2-names}


\bibliography{references.bib}  

\end{document}